\documentclass{article}
\usepackage[utf8]{inputenc}
\usepackage[affil-it]{authblk}
\usepackage{hyperref}
\usepackage{bm}
\usepackage{cite}
\usepackage{nameref}
\usepackage{float}
\usepackage{amsmath}
\usepackage{amssymb}
\usepackage{graphicx}
\usepackage{rotating}
\usepackage{caption}
\usepackage[version=3]{mhchem} 
\usepackage{comment}
\usepackage{xcolor}
\usepackage[left=4cm, right=4cm]{geometry}
\usepackage[colorinlistoftodos]{todonotes}
\usepackage[tableposition=top]{caption}

\title{Predicting Thermoelectric Transport Properties from Composition with Attention-based Deep Learning}
\author[1]{Luis M. Antunes} 
\author[1,2]{Keith T. Butler}
\author[1]{Ricardo Grau-Crespo}
\date{}
\affil[1]{\footnotesize Department of Chemistry, University of Reading, Whiteknights, Reading RG6 6DX, United Kingdom. {\normalfont l.m.antunes@pgr.reading.ac.uk}}
\affil[2]{\footnotesize School of Engineering and Materials Science Queen Mary University of London Mile End Road London E1 4NS, United Kingdom.}

\begin{document}

\maketitle

\begin{abstract}
Thermoelectric materials can be used to construct devices which recycle waste heat into electricity. However, the best known thermoelectrics are based on rare, expensive or even toxic elements, which limits their widespread adoption. To enable deployment on global scales, new classes of effective thermoelectrics are thus required. \textit{Ab initio} models of transport properties can help in the design of new thermoelectrics, but they are still too computationally expensive to be solely relied upon for high-throughput screening in the vast chemical space of all possible candidates. Here, we use models constructed with modern machine learning techniques to scan very large areas of inorganic materials space for novel thermoelectrics, using composition as an input. We employ an attention-based deep learning model, trained on data derived from \textit{ab initio} calculations, to predict a material's Seebeck coefficient, electrical conductivity, and power factor over a range of temperatures and \textit{n}- or \textit{p}-type doping levels, with surprisingly good performance given the simplicity of the input, and with significantly lower computational cost. The results of applying the model to a space of known and hypothetical binary and ternary selenides reveal several materials that may represent promising thermoelectrics. Our study establishes a protocol for composition-based prediction of thermoelectric behaviour that can be easily enhanced as more accurate theoretical or experimental databases become available.  
\end{abstract}

\section{Introduction}
Approximately 65\% to 70\% of the energy used in industrial and transportation processes is wasted as heat. \cite{rowe_kajikawa_2006} Traditional means of converting waste heat into electricity involve the use of devices such as Rankine steam engines, but these methods tend to involve machines comprised of multiple moving parts, which require maintenance and upkeep, and are difficult to scale. Thermoelectric generators, which are solid-state devices without moving parts, provide an alternative and convenient solution to waste heat recovery. \cite{snyder2008small} A thermoelectric generator is built from two semiconducting materials, one with $n$-type conductivity, and the other with $p$-type conductivity. The materials are typically assembled with electrical and thermal connections between a heat source, at temperature $T_{\mathrm{hot}}$, and a heat sink, at temperature $T_{\mathrm{cold}}$. The efficiency of a thermoelectric generator depends strongly on the temperature difference, $T_{\mathrm{hot}}$ - $T_{\mathrm{cold}}$, as well as on the physical characteristics of the materials used, which are usually summarized in the \textit{figure of merit}:
\begin{equation}
  zT = \frac{S^2 \sigma T}{\kappa}. \label{eqn:zT}
\end{equation}
Here, $S$ is the Seebeck coefficient, $\sigma$ is the electrical conductivity, $T$ is the absolute temperature, and $\kappa$ is the thermal conductivity, which contains two main contributions: the lattice thermal conductivity $\kappa_{\mathrm{latt}}$ due to crystal vibrations, and the electronic thermal conductivity $\kappa_{\mathrm{elec}}$ due to heat-carrying diffusion of electrons in the solid. The term $S^2 \sigma$ is commonly referred to as the \textit{power factor}. The higher the dimensionless figure of merit $zT$, the more efficient the thermoelectric material. Consequently, a good thermoelectric material must exhibit a large (absolute) Seebeck coefficient, good electrical conductivity, but low thermal conductivity.

Finding good thermoelectric materials with the right combination of properties is a difficult task, because of the interdependence of the properties that appear in the figure of merit. Other factors, like abundance and toxicity, further complicate the search for good candidate materials. While thermoelectricity has been a known phenomenon since the early 1800s \cite{seebeck1895magnetische, roget1832treatises}, relatively few materials have been discovered that are effective enough for practical applications. Well-studied thermoelectric materials, such as \ce{Bi2Te3} and \ce{PbTe}, are suitable for various applications, but are often too expensive or too toxic for widespread adoption. \cite{caballero2021environmentally} If thermoelectric generators are to be deployed on a scale large enough to have a positive environmental impact, new materials are needed. \cite{freer2020realising} The search for novel thermoelectrics is an active field of research. \cite{sootsman2009new, gayner2016recent, beretta2019thermoelectrics} A range of promising thermoelectric materials have been discovered experimentally, either serendipitously, or as a result of chemical intuition. Representative classes of materials that are being actively investigated include the metal chalcogenides (e.g. \ce{SnSe}, \ce{Cu2Se}) \cite{zhao2016snse, zhou2016lead, liu2012copper}, silicon-based alloys (e.g. \ce{SiGe}) \cite{dismukes1964thermal}, skutterudites (e.g. \ce{CoAs3}, \ce{CoSb3}) \cite{caillat1996bridgman}, Zintl compounds (e.g. \ce{YbZn2Sb2}) \cite{gascoin2005zintl}, clathrates (e.g. \ce{Sr8Ga16Ge30}) \cite{nolas1998semiconducting}, Heusler and Half-Heusler compounds (e.g. \ce{TiNiSn}, \ce{ZrNiSn}) \cite{aliev1989gap, aliev1990narrow, hohl1997new}, and metal oxides (e.g. \ce{NaCo2O4}, \ce{Ca3Co4O9}) \cite{terasaki1997large, tian2014enhancement}. Amongst these candidates, hole-doped polycrystalline \ce{SnSe} is the best performer in terms of thermoelectric figure of merit, and is reported to exhibit a $zT$ of 3.1 at 783K \cite{zhou2021polycrystalline}. In principle, there are no theoretical or thermodynamic limits for the possible values of $zT$ \cite{tritt2006thermoelectric}, so there is hope that materials with even higher values of $zT$ can be found.

In addition to trial-and-error exploration, and the rational design of materials, computational techniques based on the combination of density functional theory (DFT) and high-throughput screening (HTS) are becoming increasingly prevalent in the search for new thermoelectrics \cite{sparks2016data, gorai2017computationally, recatala2020toward}. The first report of such an approach was made in 2006 by Madsen, who screened a dataset of 1,630 Sb-containing compounds derived from existing crystal structure databases, and based on the results of \textit{ab initio} calculations, identified LiZnSb as an interesting thermoelectric material \cite{madsen2006automated}. Since then, a number of studies involving the use of HTS in the search for new thermoelectric candidates have followed \cite{wang2011assessing, carrete2014nanograined, toher2014high, gorai2015computational, zhu2015computational, xi2018discovery, gorai2020computational, chen2021computational, pohls2021experimental}. The increasing availability of distributed computing infrastructure, along with the development of workflow management software \cite{pizzi2016aiida, mathew2017atomate, jain2015fireworks, curtarolo2012aflow, saal2013materials, zapata2019qmflows, adorf2018simple, mayeshiba2017materials}, has enabled the growing adoption of this approach. 

 While DFT-based HTS is becoming more prevalent, there remains a large gap between the size of chemical space that is accessible with this approach, and the size of the space of all possible inorganic materials. To bridge that gap, and to further accelerate computational predictions of thermoelectric behaviour, techniques involving the use of machine learning (ML) have been gaining popularity in the search for new thermoelectric materials. \cite{wang2020machine, juneja2021accelerated, han2021machine, qian2021machine, antunes2022machine} Data for these ML approaches can come from either theoretical calculations, or from physical experiments. HTS experiments have been producing \textit{ab initio} results for thousands of materials, and these results can be assembled into datasets that are usable with ML algorithms. Since experimental data is scarcer, the outputs of \textit{ab initio} calculations are often the source of data for ML approaches. Using ML to learn models that predict the output of \textit{ab initio} calculations is sensible, since invoking an ML model is much faster (and less computationally expensive) than carrying out an \textit{ab initio} calculation. ML models of various thermoelectric properties, such as the Seebeck coefficient \cite{furmanchuk2018prediction, gaultois2016perspective, pimachev2021first, yuan2022machine}, electrical conductivity \cite{mukherjee2020statistical, yoshihama2021design}, power factor \cite{choudhary2020data, sheng2020active, yang2021accurate, laugier2018predicting}, lattice thermal conductivity \cite{carrete2014finding, seko2015prediction, zhang2018strategy, chen2019machine, juneja2019coupling, tewari2020machine, liu2020high, li2020deep, loftis2020lattice, miyazaki2021machine, tranaas2022lattice, jaafreh2021lattice, choi2022accelerated}, and even $zT$ \cite{tabib2018discovering, wang2019improved, na2021predicting, zhong2021data, gan2021prediction, jaafreh2022deep}, have been developed.

Deep learning is a particular ML approach that has been very successful in recent years, and has seen adoption in many diverse areas of science \cite{lecun2015deep, skansi2018introduction}. It is characterized by the combination of large datasets with various neural network architectures, together with advantages such as automatic feature extraction. In materials chemistry, deep learning approaches have been adopted for prediction of materials properties \cite{agrawal2019deep}. General purpose deep learning architectures for materials properties prediction, such as ElemNet \cite{jha2018elemnet}, IRNet \cite{jha2019irnet, jha2021enabling}, CGCNN \cite{xie2018crystal}, MEGNet \cite{chen2019graph}, Roost \cite{goodall2020predicting}, and CrabNet \cite{wang2021compositionally} have become powerful tools in the materials informatics toolbox.

Here, we utilize attention-based deep learning, together with existing datasets derived from high-throughput DFT calculations \cite{parr1980density}, to predict the thermoelectric transport properties of a material. The input to the model is a representation of a material's composition, and optionally the material's band gap. The output is a collection of predictions for a range of temperatures, for various doping levels, and for \textit{n} and \textit{p} doping types. This structure-free approach allows us to scan regions of materials space of hypothetical but plausible compounds, whose structures are not known. Our multi-output approach creates a thermoelectric behaviour profile for a material at a number of different conditions, which offers advantages over narrower models that only make predictions for specific conditions.

\section{Methods}

\subsection{Datasets}
Our models are trained on the dataset published in 2017 by Ricci \textit{et al.} \cite{ricci2017ab} (henceforth the \textit{Ricci database}). This is a freely available electronic transport database containing the computationally derived electronic transport properties for 47,737 inorganic compounds with stoichiometric compositions. The properties listed include the Seebeck coefficient, the electrical conductivity, and the electronic thermal conductivity, obtained using DFT in the generalized gradient approximation (GGA), and the Boltzmann Transport Equation through the BoltzTraP computer software \cite{madsen2006boltztrap}, under the constant relaxation time approximation (CRTA). They also associate the computed band gap with each entry, amongst several other properties. For each compound, the aforementioned properties were determined at various temperatures (100K to 1300K in 100K increments), for $p$- and $n$-doping types, and at 5 doping levels (ranging from $10^{16}$ to $10^{20}$ $\mathrm{cm}^{-3}$). Moreover, each property is a tensor quantity reported as a $3{\times}3$ matrix. The database is altogether quite large, with 18,617,430 data points if one considers only the values of the diagonal elements $S_{\mathrm{xx}}$, $S_{\mathrm{yy}}$, and $S_{\mathrm{zz}}$ (i.e. 47,737 compounds $\times$ 13 temperatures $\times$ 2 doping types $\times$ 5 doping levels $\times$ 3 diagonal elements). Another important consideration is that there are duplicate compounds in the database in terms of composition (corresponding to possible polymorphs). While there are 47,737 unique compounds in the database when structure is considered, there are only 34,628 unique compositions. In this study, we form a dataset of compositions from the Ricci database and their associated thermoelectric transport properties. For cases where there are multiple entries with the same composition, we obtain the DFT-derived energy per atom of each polymorph, and use the transport properties and band gap of the entry corresponding to the polymorph with the lowest energy per atom.

Additionally, we form a dataset consisting solely of compositions and their associated electronic band gaps derived from DFT, by combining data from the Materials Project \cite{jain2013commentary} and the Ricci database. We obtained 126,335 structures and their associated electronic band gaps from the Materials Project, which corresponded to 89,444 unique compositions, which are used to train the band gap predictor. Where there were multiple structures for a composition, again we used the band gap of the polymorph with the lowest computed energy per atom.

The Ricci database has some important limitations. As discussed in Ref. \cite{ricci2017ab} and elsewhere (see Ref. \cite{plata2022silico} for a recent perspective), the use of the GGA and CRTA in the prediction of electronic transport can lead to large discrepancies with respect to experiment. In particular, GGA band structures generally exhibit too narrow gaps and too large bandwidths, which tends to exaggerate the electronic conductivity. The CRTA, especially when unaccompanied by physically-sound prediction of relaxation times, misses important differences in scattering mechanisms across compounds. Inevitably, any ML model based on this dataset will carry over these limitations of the underlying data, hindering the quality of the predictions with respect to experimental values. However, our approach establishes a protocol capable of efficiently mapping composition to thermoelectric behavior, which can be easily refined once more accurate databases become available. This is important because, in addition to the improvement of existing \textit{ab initio} databases, there are ongoing efforts to create large databases of thermoelectric properties from experiments \cite{freer2022key}, so we anticipate our model will keep evolving following the expansion of such datasets.

\subsection{ML Models}

We build ML models that predict the Seebeck coefficient, the electrical conductivity, and the power factor using data from the Ricci database. Our multi-output regression models \cite{borchani2015survey, xu2019survey} produce predictions of transport properties at 13 temperatures, 5 doping levels, for 2 doping types, given a material's composition and (optionally) band gap. The task is to predict the mean of the diagonal elements of the Seebeck tensor, $(S_{\mathrm{xx}}+S_{\mathrm{yy}}+S_{\mathrm{zz}})/3$, henceforth referred to as the Seebeck coefficient, $S$, and the mean of the diagonal elements of the electrical conductivity tensor, $(\sigma_{\mathrm{xx}} + \sigma_{\mathrm{yy}} + \sigma_{\mathrm{zz}})/3$, henceforth referred to as the electrical conductivity, $\sigma$. The values for electrical conductivity in the Ricci database are reported per unit of relaxation time. Hence, in this report, electrical conductivity, $\sigma$, will more precisely refer to electrical conductivity per unit relaxation time, $\sigma/\tau$. The target power factor, $S^2\sigma$, is also predicted, and is defined here as the mean of the directional power factors, $(S_{\mathrm{xx}}^2\sigma_{\mathrm{xx}}+S_{\mathrm{yy}}^2\sigma_{\mathrm{yy}}+S_{\mathrm{zz}}^2\sigma_{\mathrm{zz}})/3$. It will be denoted by $PF$.

More formally, the task is to learn a function $f:\mathcal{X} \to \mathcal{Y}$, given a training set $\mathcal{D} = \{ (\mathbf{x}_i, \mathbf{y}_i) \mid 1 \leq i \leq k \}$, with $\mathbf{x}_i \in \mathcal{X}$, $\mathbf{y}_i \in \mathcal{Y}$, and $k$ labeled examples. Here, the $\mathbf{x}_i$ represent a multi-dimensional input describing the features of an exemplar, and $\mathbf{y}_i$ represent a multi-dimensional target associated with $\mathbf{x}_i$. A training procedure is used to find $f$, and involves the minimization of a loss, $L:\mathcal{Y} \times \hat{\mathcal{Y}} \to \mathbb{R}$, that specifies the degree of disagreement between the true values $\mathcal{Y}$, and $\hat{\mathcal{Y}}$, the output of $f$ given members of $\mathcal{X}$.

Here, we primarily use two different forms of $f$: a Random Forest \cite{ho1995random}, and an attention-based deep neural network based on the CrabNet architecture \cite{wang2021compositionally}. The CrabNet architecture incorporates a multi-head self-attention mechanism, originally introduced in the Transformer deep learning model \cite{vaswani2017attention}. Traditionally, a Transformer transforms an input sequence to an output sequence using an encoder followed by a decoder. However, CrabNet consists strictly of an encoder, followed by a number of Residual blocks \cite{he2016deep}. Moreover, instead of a sequence of words, CrabNet operates on a bag of atoms, and consequently, instead of using a positional encoding of the input, it encodes the relative amounts of atoms present.

\begin{figure}[H]
	\includegraphics[scale=0.40]{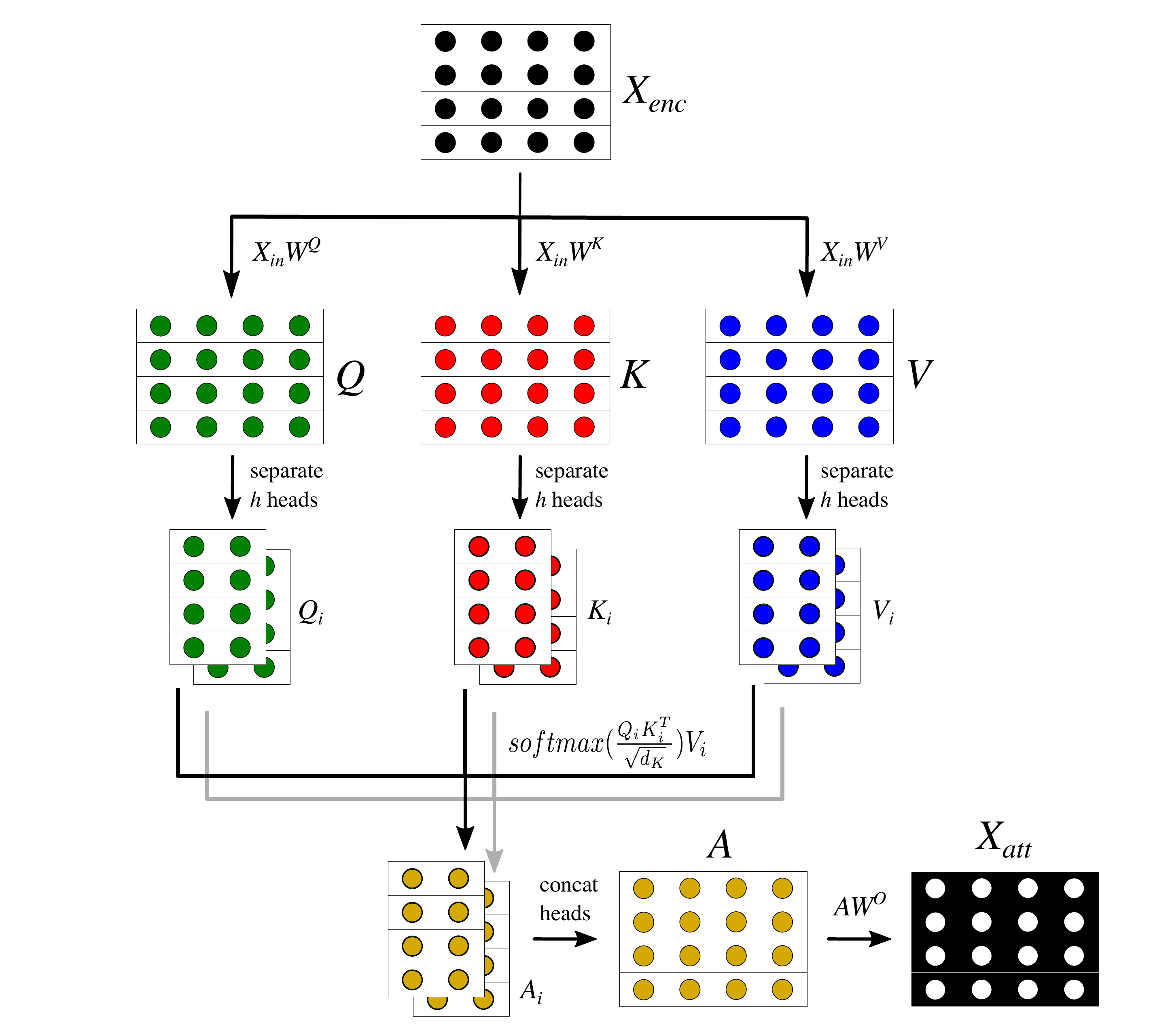}
	\caption{Depiction of the multi-head self-attention operation. The input, $X_{\mathrm{enc}} \in \mathbb{R}^{n \times d_{\mathrm{model}}}$, consists of the fractionally encoded representations for $n$ constituent elements of the composition, each with $d_{\mathrm{model}}$ components. Linear transformations are applied to the input to produce the query, $Q \in \mathbb{R}^{n \times d_{\mathrm{model}}}$, the key, $K \in \mathbb{R}^{n \times d_{\mathrm{model}}}$, and the value, $V \in \mathbb{R}^{n \times d_{\mathrm{model}}}$, using the learned parameters $W^Q$, $W^K$, and $W^V$, respectively. The query, key and value are each subsequently separated into $h$ heads (indexed here by $i$). The corresponding queries, $Q_i$, keys, $K_i$, and values, $V_i$, are combined to produce the attention products, $A_i$, by multiplying the $softmax$ of a scaled dot-product of the queries and keys with the values. After the attention products are concatenated to produce $A$, a linear transformation of $A$ using the learned parameters $W^O$ produces the output of multi-head self-attention, $X_{att} \in \mathbb{R}^{n \times d_{\mathrm{model}}}$.}
	\label{fgr:selfattention}
\end{figure}

The input to the model thus consists of a material's composition. Formally, the input, $X_{\mathrm{in}} \in \mathbb{R}^{n \times d_{\mathrm{in}}}$, consists of $d_{\mathrm{in}}$-dimensional representations for the $n$ constituent elements of the composition. The first step involves the encoding of the relative amounts of atoms into $X_{\mathrm{in}}$, referred to as \textit{fractional encoding} (see \cite{wang2021compositionally} for more details), resulting in $X_{\mathrm{enc}} \in \mathbb{R}^{n \times d_{\mathrm{model}}}$, where $d_{\mathrm{model}}$ is given as a hyperparameter. This is followed by the sequential application of a number of Transformer blocks. Each Transformer block begins by performing a multi-head self-attention operation. (Figure \ref{fgr:selfattention}) The self-attention operation allows the model to learn to attend to the relationships between the atoms of the composition, in the context of the task. The ``attention weights'' are encoded into a $n \times n$ matrix, associated with each of $h$ attention heads, by applying the $softmax$ operation to a scaled dot-product of a query, $Q_i \in \mathbb{R}^{n \times d_K}$, and a transposed key, $K_i^T \in \mathbb{R}^{d_K \times n}$, where $d_K = d_{\mathrm{model}}/h$ specifies the key (and query) dimension for an attention head.

\begin{figure}[H]
    \centering
	\includegraphics[scale=0.55]{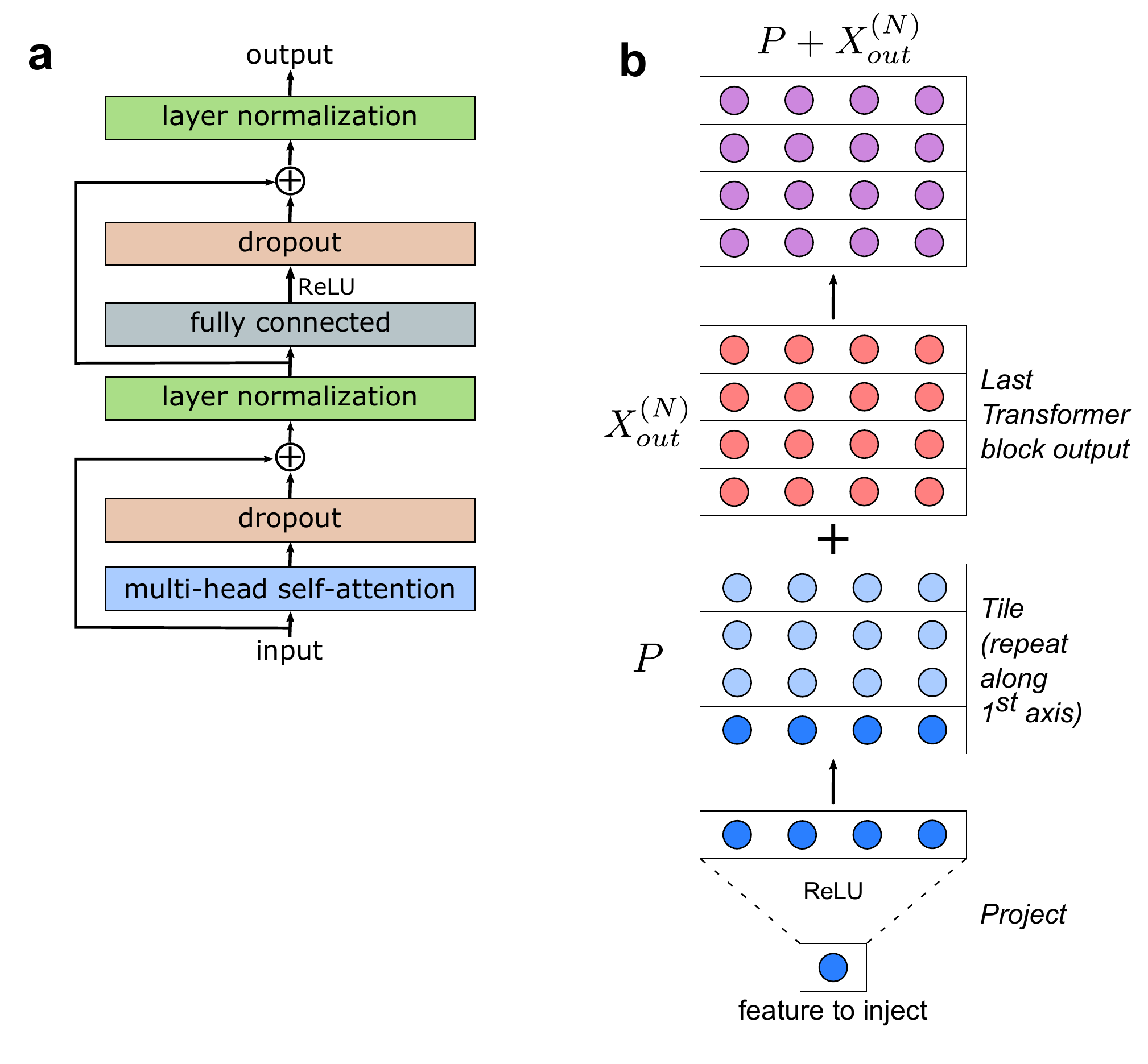}
	\caption{\textbf{a}) Components of the Transformer block. \textbf{b}) A depiction of how additional features, such as band gap, are incorporated into the CrabNet architecture. The addition operation refers to element-wise addition. The projection of the feature involves trainable parameters.}
	\label{fgr:block-and-inject}
\end{figure}

The Transformer block follows the multi-head self-attention operation with layer normalization \cite{ba2016layer}, dropout \cite{srivastava2014dropout}, and feed-forward $ReLU$ operations (Figure \ref{fgr:block-and-inject}a). The output of a Transformer block, $X_{\mathrm{out}} \in \mathbb{R}^{n \times d_{\mathrm{model}}}$, thus consists of the same dimensions as the input, which allows multiple Transformer blocks to be connected serially.

Since it may also be desirable to provide additional information beyond  composition to the model, we augment the CrabNet architecture so that additional features may be provided. There are a number of ways this could be accomplished, but we choose to borrow an approach from computer vision \cite{levine2018learning}, and perform a projection on $v$ input features, $\mathbf{u} \in \mathbb{R}^v$, followed by a tiling operation, so that the resulting projected features, $P \in \mathbb{R}^{n \times d_{\mathrm{model}}}$, have the same dimensions as the output of a Transformer block. Finally, we perform element-wise addition, $P + X_{\mathrm{out}}^{(N)}$, where $N$ is the number of Transformer blocks, and $X_{\mathrm{out}}^{(N)}$ denotes the output of the last Transformer block (Figure \ref{fgr:block-and-inject}b). While any number of extra features may be supplied to the model this way, in this work, we (optionally) supply a single feature, the band gap $E_\mathrm{g}$, associated with the material.

\begin{figure}[H]
    \centering
	\includegraphics[scale=0.8]{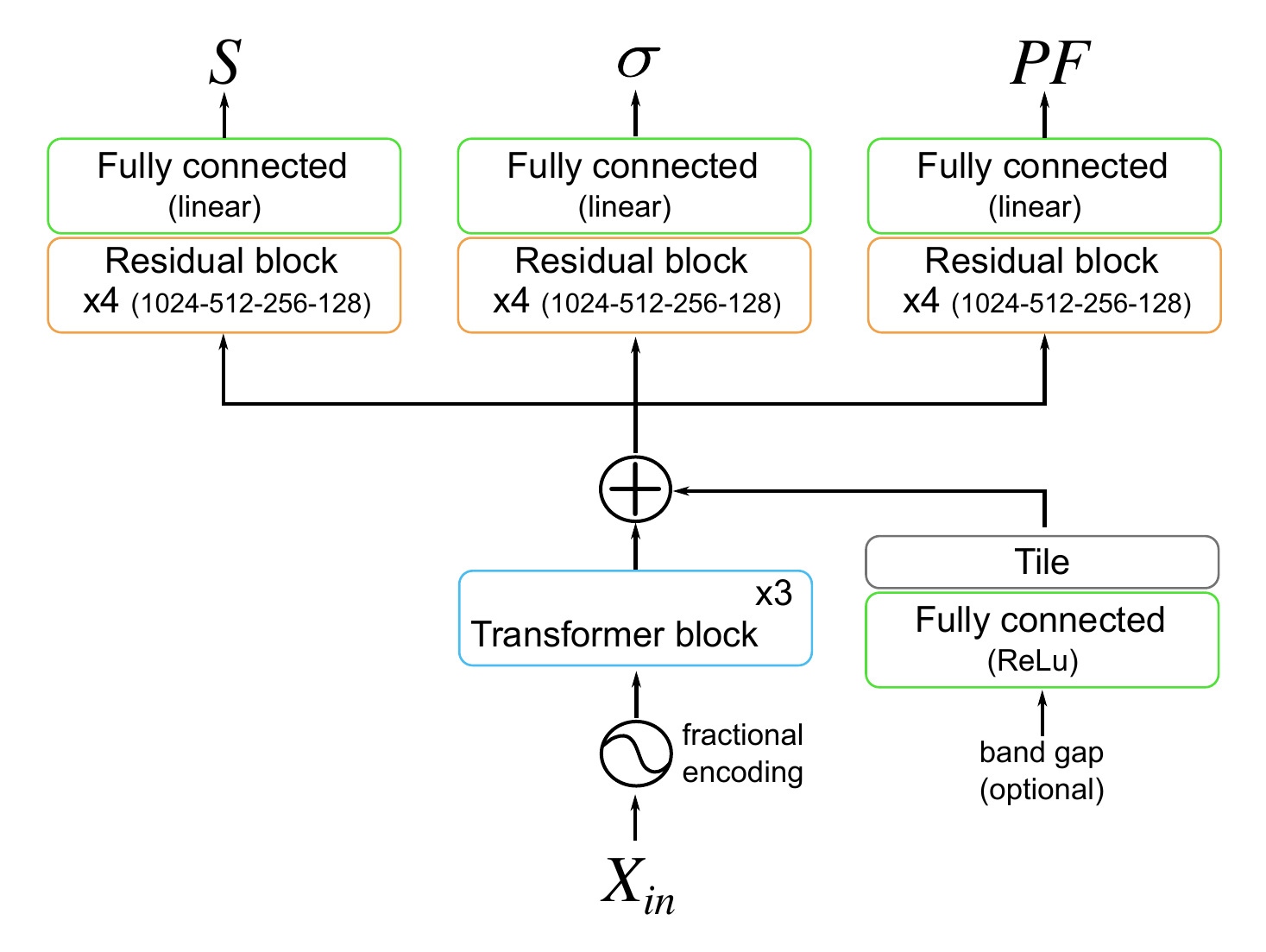}
	\caption{The multi-head attention-based architecture, CraTENet, used in this study. Each of the three output heads are multi-valued, containing the prediction of the Seebeck ($S$), electrical conductivity ($\sigma$), and power factor ($PF$), at different temperatures, doping levels, and doping types.}
	\label{fgr:architecture}
\end{figure}

Finally, the output $P + X_{\mathrm{out}}^{(N)}$ is given to three separate output heads. Each output head consists of a series of Residual blocks, followed by a fully connected linear layer that produces the final predictions for each of $S$, $\sigma$, and $PF$. This multi-head architecture has advantages in terms of convenience, efficiency, and also usually provides better overall performance on the task when compared to using a separate (single-head) model for each property predicted. (See Supplementary Table 1 for a comparison of the performance of architectures with different output head numbers.) For clarity, and to differentiate it from the original CrabNet architecture, we refer to this model as CraTENet (Compositionally-restricted attention-based ThermoElectrically-oriented Network); its architecture is illustrated in Figure \ref{fgr:architecture}.

The CraTENet model thus expects a dataset consisting of compositions, $X_i \in \mathbb{R}^{n \times d_{\mathrm{in}}}$, and associated thermoelectric transport properties, $\mathbf{y}_i^{S}, \mathbf{y}_i^{\sigma}, \mathbf{y}_i^{PF} \in \mathbb{R}^m$, where $\mathbf{y}_i^{S}$, $\mathbf{y}_i^{\sigma}$, and $\mathbf{y}_i^{PF}$, represent the $S$, $\sigma$, and $PF$ transport values, respectively, at all temperatures, doping levels and doping types, each an $m$-dimensional vector. Optionally, a band gap, $E_{\mathrm{{g}_i}} \in \mathbb{R}$, may be associated with $X_i$. The dataset is thus $\{ ((X_i, E_{\mathrm{g}_i}), (\mathbf{y}_i^{S}, \mathbf{y}_i^{\sigma}, \mathbf{y}_i^{PF})) \mid 1 \leq i \leq k \}$, where $k$ is the number of examples.

As in the CrabNet and Roost models, the CraTENet model learns the heteroscedastic aleatoric uncertainty (\textit{i.e.} how the variance of the predicted variable depends on the independent variables), explicitly through the loss function \cite{nix1994estimating, kendall2017uncertainties}. Here, the calculated variance is a measure of the uncertainty associated with the incompleteness of the descriptor used (which is why the calculated variance decreases considerably when the band gap information is added to the descriptor). This variance is different from the epistemic variance related to the quality of the model parameterization.  Whereas the CrabNet and Roost models use a Robust L1 loss to estimate the uncertainty, we find that a Robust L2 loss, which places an L2 distance on the residuals, results in superior performance for this task (see Supplementary Note 2 and Supplementary Table 4). The loss, $L_p$, for a particular thermoelectric transport property $p$, is given by:

\begin{equation}
    L_p = \frac{1}{2k}\sum_{i=1}^{k} \sum_{j=1}^{m} (\hat{y}_{ij}^p - y_{ij}^p)^{2} \exp({-\ln \hat{s}_{ij}^p}) + \ln \hat{s}_{ij}^p
\end{equation}

where $k$ is the number of examples in the dataset, and $m$ is the number of components of the output vector $\mathbf{y}_i^p$. The prediction of the $i^{\mathrm{th}}$ example is $\hat{\mathbf{y}}_i^p$, and $\hat{y}_{ij}^p$ the $j^{\mathrm{th}}$ component of the $i^{\mathrm{th}}$ prediction (also considered the predictive mean in this context). The corresponding target value is $y_{ij}^p$. Finally, the predictive aleatoric variance for the $j^{\mathrm{th}}$ component of the $i^{\mathrm{th}}$ prediction is given by $\hat{s}_{ij}^p$. The form of this loss arises from the assumption that the uncertainty in the observations follows a Gaussian distribution. Also, the term $\exp({-\ln \hat{s}_{ij}^p})$ is used in place of the term $1 / \hat{s}_{ij}^p$ for numerical stability reasons, such as to avoid a potential division by zero. Since the model utilizes a separate output head for each of the three thermoelectric transport properties being learned, the overall loss, $L$, to be minimized is given by:
\begin{equation}
    L = \alpha L_{S} + \beta L_{\sigma} + \gamma L_{PF}
\end{equation}
where $\alpha$, $\beta$, and $\gamma$ are constants which weight the importance of each of the terms in the loss $L$. In this work, $\alpha = \beta = \gamma = 1$.

Finally, we also train a band gap predictor from composition, using the original CrabNet model and the expanded band gap dataset described previously. The fact that the band gap predictor can be trained with a much larger dataset than the one used for training the CraTENet model justifies our attempt to use the band gap as an additional input to CraTENet for the prediction of transport coefficients. As discussed in the \nameref{discussion}, if the band gap predictor is sufficiently accurate, the inclusion of the predicted band gap in the CraTENet input can lead to overall performance enhancement, even if composition remains the only global input of the model. 

\subsection{ML Model Training and Evaluation}

For all CraTENet and CrabNet models, the input, $X_{\mathrm{in}}$, consisted of $n=8$ elements, and was zero-padded if the composition consisted of less than 8 elements. Each element in the input was described with a SkipAtom distributed representation \cite{antunes2022distributed} with dimensions $d_{\mathrm{in}} = 200$. (We performed experiments, as described in Supplementary Note 1 and Supplementary Table 3, to determine the performance of different descriptors). The default architectural hyperparameters of the original CrabNet model were used without further tuning. Specifically, both models consisted of $h=4$ attention heads in each of 3 sequential Transformer blocks; the hyperparameter $d_{\mathrm{model}}$ was set to 512. The output (or output head) consisted of 4 sequential Residual blocks, with 1024, 512, 256, and 128 nodes respectively. For all neural network training procedures, a mini-batch size of 128 and a learning rate of $10^{-4}$ was used, which were derived from a hyperparameter grid search. The Adam optimizer, with an epsilon parameter of $10^{-8}$, was used. \cite{kingma2014adam} All neural network models were implemented using the TensorFlow \cite{abadi2016tensorflow} and Keras \cite{chollet2015keras} software libraries. 

The input for the Random Forest models was a descriptor described by Meredig \textit{et al.} \cite{meredig2014combinatorial}, as implemented in the Matminer software library \cite{ward2018matminer}. It is a local descriptor of composition, containing properties such as atomic fractions, electronegativities, and radii. In some experiments, we concatenate an unscaled band gap feature to the descriptor. The random forest model hyperparameters were determined using a grid search. The number of estimators was set to 200, the maximum depth was set to 110, the maximum number of features was set to 36, and bootstrapping was used. We used the implementation provided in the Scikit-learn software library \cite{scikit-learn}.

Because the electrical conductivity values in the Ricci database are given per unit of relaxation time $\tau$, which is an exceedingly small number (i.e. $10^{-15}$ s), the target values for $\sigma$ and $PF$ are numerically quite large. The values also vary by orders of magnitude, reflecting the distribution across metal, semiconductor and insulating conductivity ranges. For these reasons, the models learn $\log_{10} \sigma$ and $\log_{10} PF$ instead. All output targets are standardized by removing the mean and scaling to unit variance. The band gap, when it is provided to the CraTENet model, is given in eV units and unscaled.

Neural network model training was carried out in one of two contexts: a 90-10 holdout experiment, or a 10-fold cross-validation experiment. For 90-10 holdout experiments, we split the dataset $\mathcal{D}$ into a set $\mathcal{A}$ consisting of 90\% of the data, and a set $\mathcal{B}$ consisting of 10\% of the data. For the neural network models, set $\mathcal{A}$ was further split into a training set $\mathcal{T}$ consisting of 90\% of $\mathcal{A}$, and a validation set $\mathcal{V}$ consisting of 10\% of $\mathcal{A}$. Early stopping was used (with a patience of 50) to determine the optimal number of epochs to train, using $\mathcal{V}$ as the validation set. Then, the model was re-trained on all of $\mathcal{A}$ for the number of epochs determined to be optimal, again starting from random parameters. Test set $\mathcal{B}$ was then used to evaluate performance of the re-trained model (see \cite{goodfellow2016deep} for more information on this approach). The Random Forest models were trained on $\mathcal{A}$, and evaluated on $\mathcal{B}$. The same random seed was used throughout when creating the splits, to ensure identical splits for all experiments.

For the 10-fold cross-validation experiments, we followed the same procedure as for the 90-10 holdout experiments, except that we create 10 mutually exclusive splits, each consisting of 10\% of $\mathcal{D}$ for testing and 90\% of $\mathcal{D}$ for training, using the same random seed for all experiments, and repeating the hold-out procedure for each of the 10 splits. The performance on $\mathcal{B}$ was averaged across the 10 splits to yield the final performance of the model.

The objective of all neural network training experiments was to minimize either the Robust L1 or Robust L2 loss. The objective of Random Forest training was to minimize the mean squared error (MSE) criterion. The mean absolute error (MAE) and coefficient of determination ($R^2$) metrics were used to assess model performance. To produce the final neural network models to be used for inference on composition space outside the datasets used for training and evaluation, we train the models on all available data $\mathcal{D}$ for a number of epochs determined from the corresponding 10-fold cross-validation experiment, by averaging the number of epochs required for each fold. The final Random Forest models to be used for inference were simply trained on all available data $\mathcal{D}$.

\subsection{DFT Calculations}

We performed a small number of DFT calculations in systems not found in the Ricci database, for testing purposes. All calculations were carried out using the Vienna Ab initio Simulation Package (VASP) \cite{kresse1993ab, kresse1996efficient}, and the calculation settings were chosen to follow the work of Ricci \textit{et al.} \cite{ricci2017ab} as closely as possible. The Perdew-Burke-Ernzerhof (PBE) \cite{perdew1996generalized} exchange-correlation functional, which is based on the GGA, was used in conjunction with the projector augmented-wave method \cite{blochl1994projector, kresse1999ultrasoft} to describe the interaction between core and valence electrons. All structures were fully relaxed until the force on each atom is below 0.02 eV/\r{A}. Spin polarization was on, and magnetic moments on the ions were initialized in a high-spin ferromagnetic configuration, and then allowed to relax to the spin groundstate. A self-consistent static calculation was performed using 90 $k$-points/$\text{\r{A}}^{-3}$ (in terms of reciprocal lattice volume) for systems with band gaps $\ge 0.5$ eV, and 450 $k$-points/$\text{\r{A}}^{-3}$ for systems with band gaps $< 0.5$ eV. Subsequently, a non-self-consistent calculation was performed to evaluate the band structures on a uniform $k$-point grid, with 1,000 $k$-points/$\text{\r{A}}^{-3}$ for systems with band gaps $\ge 0.5$ eV, and 1,500 $k$-points/$\text{\r{A}}^{-3}$ for systems with band gaps $< 0.5$ eV. Spin-orbit coupling was not considered. 

The Seebeck coefficient, $S$, and the electrical conductivity, $\sigma$, were computed using the BoltzTraP2 software package \cite{BoltzTraP2}. Interpolation was first performed by sampling 5 irreducible $k$-points for each $k$-point from the VASP output. The band structure was then integrated to obtain sets of Onsager coefficients. The temperature range 100K to 1300K was explored, in increments of 100K, at 5 different doping levels ($10^{16}$ to $10^{20}$ $\mathrm{cm}^{-3}$), for both $n$ and $p$ doping types.

\section{Results and Discussion} \label{discussion}

\subsection{Thermoelectric Property Prediction}

Both the CraTENet model and a Random Forest model were trained on the 34,628 entries of the Ricci database. To establish the generalization error of the models, 10-fold cross-validation was performed. Since multi-target regression of thermoelectric transport properties on composition is essentially a new task, unreported in the literature, there are no existing benchmarks to compare with. We created simple baseline models, such as linear regression with a Meredig feature vector, or simply taking the median of the target values, but these models performed considerably worse than the ML models presented here. To simplify presentation, we leave out the baseline results.

The results of 10-fold cross-validation are presented in Table \ref{tab:ten-fold-cv-results}. For the remainder of this article, ``CraTENet'' will refer to either the version of the model which does not accept a band gap input or to the CraTENet model in general, depending on the context, whereas ``CraTENet+gap'' will specifically refer to the version of the model which requires a band gap input. As is evident from the results in Table \ref{tab:ten-fold-cv-results}, the models which utilize the band gap clearly outperform those which do not. The band gap is thus an important predictor of thermoelectric transport properties. In both the case where band gap is or is not provided, the CraTENet model outperforms the Random Forest model in terms of MAE. The Random Forest performs better in terms of $R^2$, but generally only when band gap is absent. Moreover, the models appear to perform slightly better when predicting the $\log \sigma$ than the Seebeck. Prediction of the $\log PF$ appears to be the most problematic, with the $R^2$ for this property being noticeably lower than for the other two properties. The best thermoelectric materials have Seebeck coefficients in the order of several hundreds of $\mu V/K$, so the resulting MAE is still reasonably small by comparison.

\begin{table}[H]
\centering
\caption {Ten-fold cross-validation results for each of the transport properties for the CraTENet and Random Forest (RF) models, both with and without a provided band gap, in terms of MAE and $R^2$. Each value represents the mean result across 10 folds, across all temperatures, doping levels and doping types. Bold values represent the best result for a class of models (i.e. with or without band gap) for a particular property.}
\begin{tabular}{ c c c c c c c } 
\hline
& \multicolumn{2}{c}{$S$} & \multicolumn{2}{c}{$\log \sigma$} & \multicolumn{2}{c}{$\log PF$} \\
& MAE ($\mu$V/K) & $R^2$ & MAE & \bf $R^2$ & MAE & $R^2$  \\
\hline
CraTENet & \bf 114 & 0.780 & \bf 0.576 & 0.768 & \bf 0.452 & 0.616 \\
RF & 141 & \bf 0.798 & 0.696 & \bf 0.780 & 0.476 & \bf 0.632 \\
\hline
CraTENet+gap & \bf 49 & \bf 0.962 & \bf 0.260 & \bf 0.968 & \bf 0.380 & 0.731 \\
RF+gap & 54 & 0.961 & 0.301 & 0.964 & 0.398 & \bf 0.737 \\
\hline
\end{tabular}
\label{tab:ten-fold-cv-results} 
\end{table}

\begin{figure}[H]
    \centering
    \includegraphics[scale=0.69]{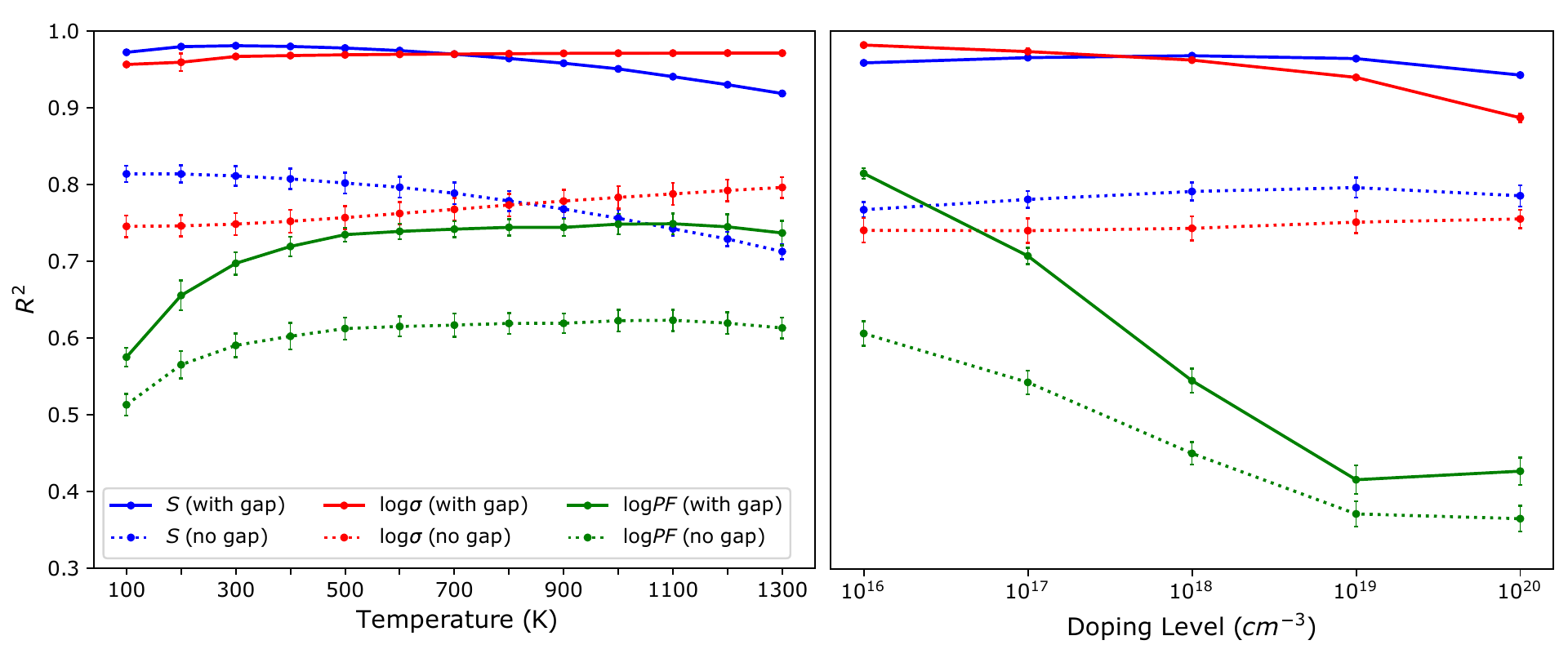}
	\caption{Ten-fold cross-validation performance ($R^2$) of the CraTENet model as a function of temperature and doping level. On the left, each point represents the mean performance for each temperature across all 10 folds, across all doping levels and doping types. On the right, each point represents the mean performance for each doping level across all 10 folds, across all temperatures and doping types. The dotted series represent the model's results without a provided band gap.}
	\label{fgr:temp-dop-results-v2}
\end{figure}

The results in Table \ref{tab:ten-fold-cv-results} represent predictions made for all temperatures, doping levels and doping types. However, it is useful to understand how the models perform for different cross-sections of the data. For example, the 10-fold cross-validation results as a function of doping type are presented in Table \ref{tab:ten-fold-cv-dop-type-results}. To obtain the values in Table \ref{tab:ten-fold-cv-dop-type-results}, only the predictions for a given doping type were considered when computing the metrics, across all doping levels and temperatures. The CraTENet model appears to perform better on the $p$-type predictions, though it depends on which metric one considers. In Figure \ref{fgr:temp-dop-results-v2}, 10-fold cross-validation results are presented as a function of temperature and doping level. It is interesting (and useful to know) that the $PF$ predictions are worse at lower temperatures and higher doping levels.

\begin{table}[H]
\centering
\caption {Ten-fold cross-validation performance of the CraTENet model as a function of doping type. Each value represents the mean result for each doping type across all 10 folds, across all temperatures and doping levels. Bold values represent the best result between $p$- and $n$-doping types for a class of models (i.e. with or without band gap) for a particular property.}
\begin{tabular}{ c c c c c c c c } 
\hline
& & \multicolumn{2}{c}{$S$} & \multicolumn{2}{c}{$\log \sigma$} & \multicolumn{2}{c}{$\log PF$} \\
& Doping & MAE ($\mu$V/K) & $R^2$ & MAE & \bf $R^2$ & MAE & $R^2$  \\
\hline
CraTENet & $p$-type & 119 & \bf 0.636 & 0.589 & \bf 0.775 & 0.465 & \bf 0.631 \\
CraTENet & $n$-type & \bf 109 & 0.627 & \bf0.562 & 0.758 & \bf 0.439 & 0.594 \\
\hline
CraTENet+gap & $p$-type & \bf 49 & \bf 0.945 & 0.260 & \bf 0.972 & 0.388 & \bf 0.747 \\
CraTENet+gap & $n$-type & 50 & 0.925 & 0.260 & 0.962 & \bf 0.371 & 0.709 \\
\hline
\end{tabular}
\label{tab:ten-fold-cv-dop-type-results} 
\end{table}

\begin{figure}[H]
    \centering
    \includegraphics[scale=0.65]{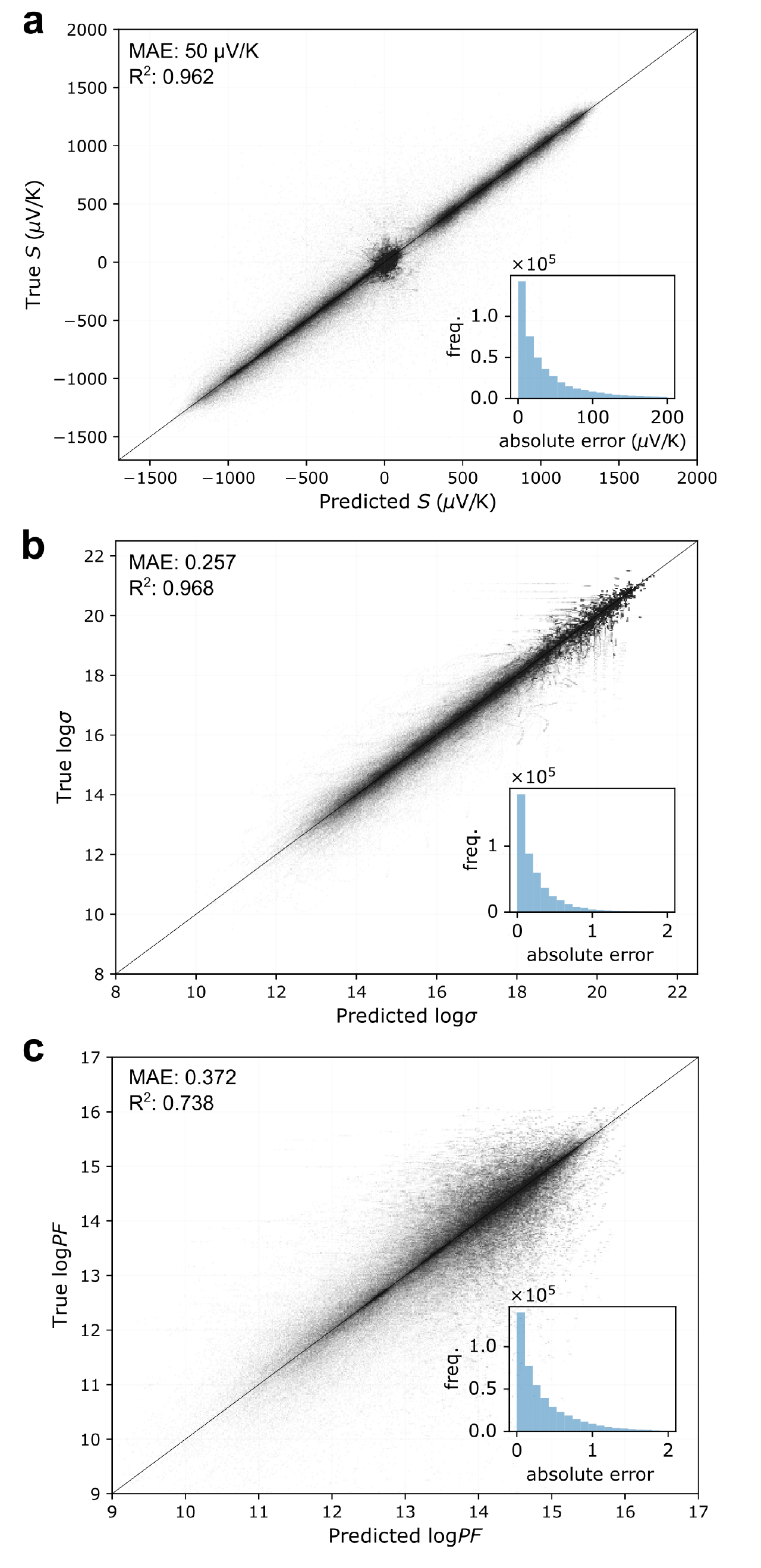}
	\caption{True values vs. predicted values of the test set of a 90-10 holdout experiment using the CraTENet+gap model, for each of the transport properties, across all temperatures, doping levels, and doping types. Each plot contains 450,190 points, as there are 3,463 compositions in the test set, each with 130 (13 temperatures $\times$ 5 doping levels $\times$ 2 doping types) associated values. The inset plots depict the distribution of absolute errors.}
	\label{fgr:true-vs-pred-props}
\end{figure}

To understand how the predictions compare to the ``true'' values (\textit{i.e.} the target DFT values), and how the prediction errors are distributed, it is useful to plot the true versus the predicted values, and also the distribution of absolute errors, as in Figure \ref{fgr:true-vs-pred-props}. As suggested by the $R^2$ values, the plots show that most predictions lie close to the true values. Moreover, the distribution of absolute errors indicates that the majority of errors are less than the overall MAE values.

\begin{figure}[h]
    \centering
    \includegraphics[scale=0.7]{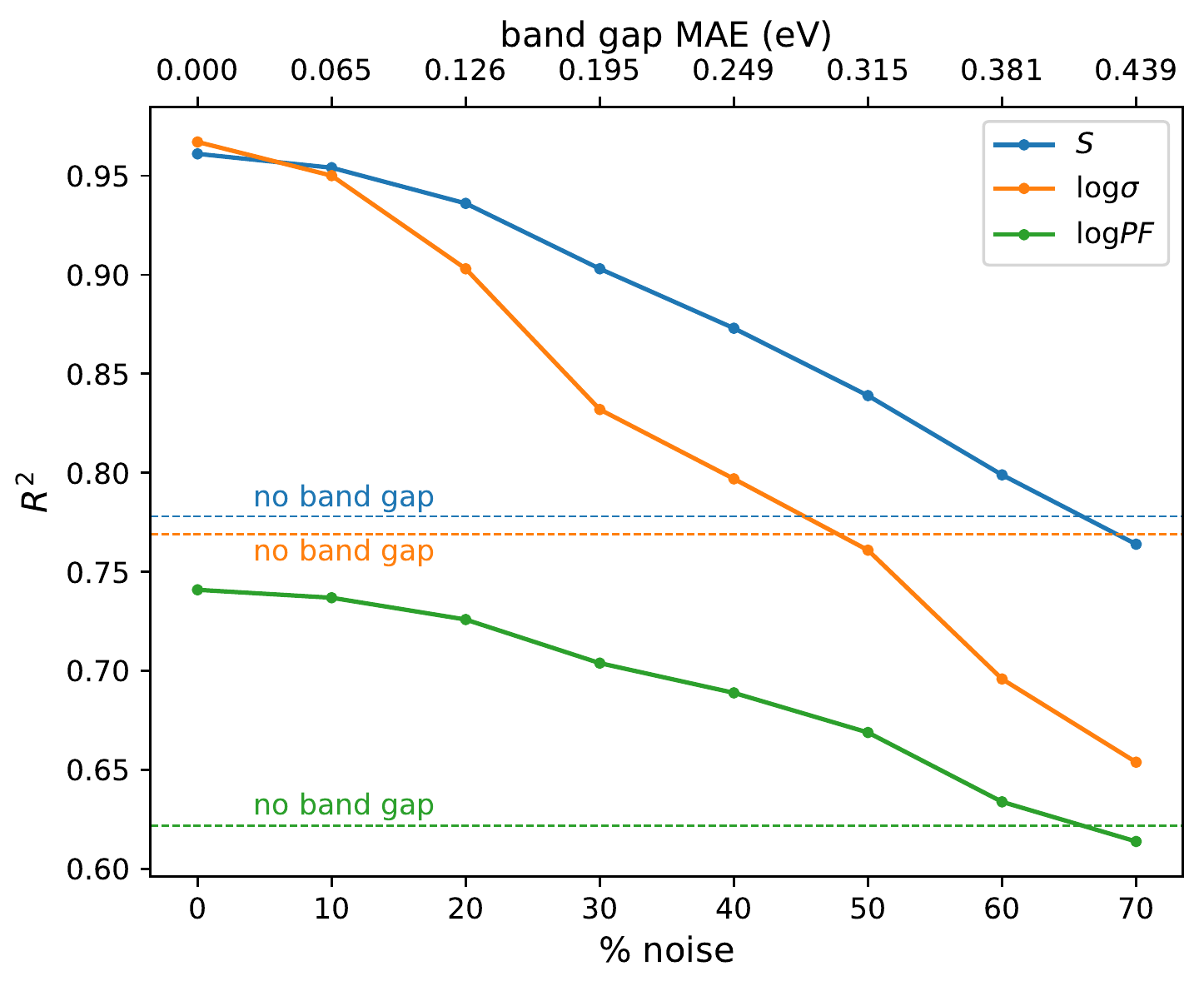}
	\caption{Performance of the CraTENet+gap model (in terms of $R^2$) as a function of band gap quality. A 90-10 holdout experiment was performed, and the actual gaps in the test set were corrupted by adding increasing amounts of Gaussian noise, before the performance of the model was assessed. The dotted lines represent the performance of the CraTENet model without a provided band gap.}
	\label{fgr:gap-sens}
\end{figure}

As the CraTENet model performs best when access to a band gap is available, it is important to understand how the performance of the model depends on the quality of the band gap provided, since, in many contexts, an experimental or \textit{ab initio} band gap may not be available. In screening scenarios, the band gap could originate from a predictive model. Thus, to understand how the CraTENet model depends on the quality of the band gap, we performed sensitivity experiments, by incrementally degrading high quality band gaps (i.e. derived from \textit{an initio} methods) by adding Gaussian noise, and then supplying these ``lower-quality'' band gaps to the model. The results are presented in Figure \ref{fgr:gap-sens}. In the figure, the horizontal axis along the top of the plot represents the resulting MAE (in eV) after a certain percentage of noise has been added to the band gaps. For example, when 10\% noise has been added to the \textit{ab initio} band gaps, the MAE when comparing these corrupted gaps to the true gaps is 0.065 eV. Figure \ref{fgr:gap-sens} shows, as might be expected, that when more noise is added to the band gaps, the performance of the model falls. However, some thermoelectric transport properties are more robust (or more sensitive) to changes in the band gap quality. For example, in the case of the prediction of the Seebeck, even with band gaps exhibiting an MAE of 0.30 eV, the model is still able to achieve an $R^2$ of 0.85, in comparison to an $R^2$ of below 0.80 when no band gap is provided. However, in the case of $\log \sigma$, the model is much more sensitive. Current state-of-the-art band gap predictors that operate on composition alone typically achieve an MAE of 0.30-0.45 eV. \cite{wu2020machine} However, band gap predictor performance is expected to improve over time, and this will further increase the utility of the CraTENet model in screening scenarios with predicted band gaps.

\subsection{Band Gap Prediction}

A dataset consisting of compositions and their corresponding DFT-derived band gaps was formed by taking all of the unique compositions in the Materials Project, and consisted of 89,444 entries. A CrabNet model was trained on this dataset, using the minimization of the Robust L1 loss as the objective. To establish the generalization error of the model, 10-fold cross-validation was performed (as described in the Methods). Across the 10 folds, the model achieved a mean $R^2$ of 0.71, and a mean MAE of 0.38 eV. A final model was trained on all 89,444 entries for 101 epochs, which was determined to be the ideal number of epochs required (\textit{i.e.} the mean number of epochs required across the 10 folds). This band gap predictor was subsequently used to provide band gaps when scanning composition space where structure and band gaps were unknown.

\subsection{Searching Composition Space for New Thermoelectrics}

\subsubsection{Materials Project Compounds not in the Ricci Database}

Of the 126,335 structures we obtained from the Materials Project, we derived 89,444 unique compositions. Since the compounds in the Ricci database originate from the Materials Project, we obtained 54,816 unique compositions when removing the compositions found in the Ricci database. This collection of 54,816 compositions forms a sizeable and convenient search space, since GGA band gaps have already been computed for these compounds, and their structures are known. Thus, we apply our CraTENet+gap model to this space, in an attempt to surface novel compounds which may represent promising thermoelectrics. We verify the quality of our predictions by performing \textit{ab initio} calculations for a small subset of these compounds.

\begin{figure}[H]
    \centering
	\includegraphics[scale=0.9]{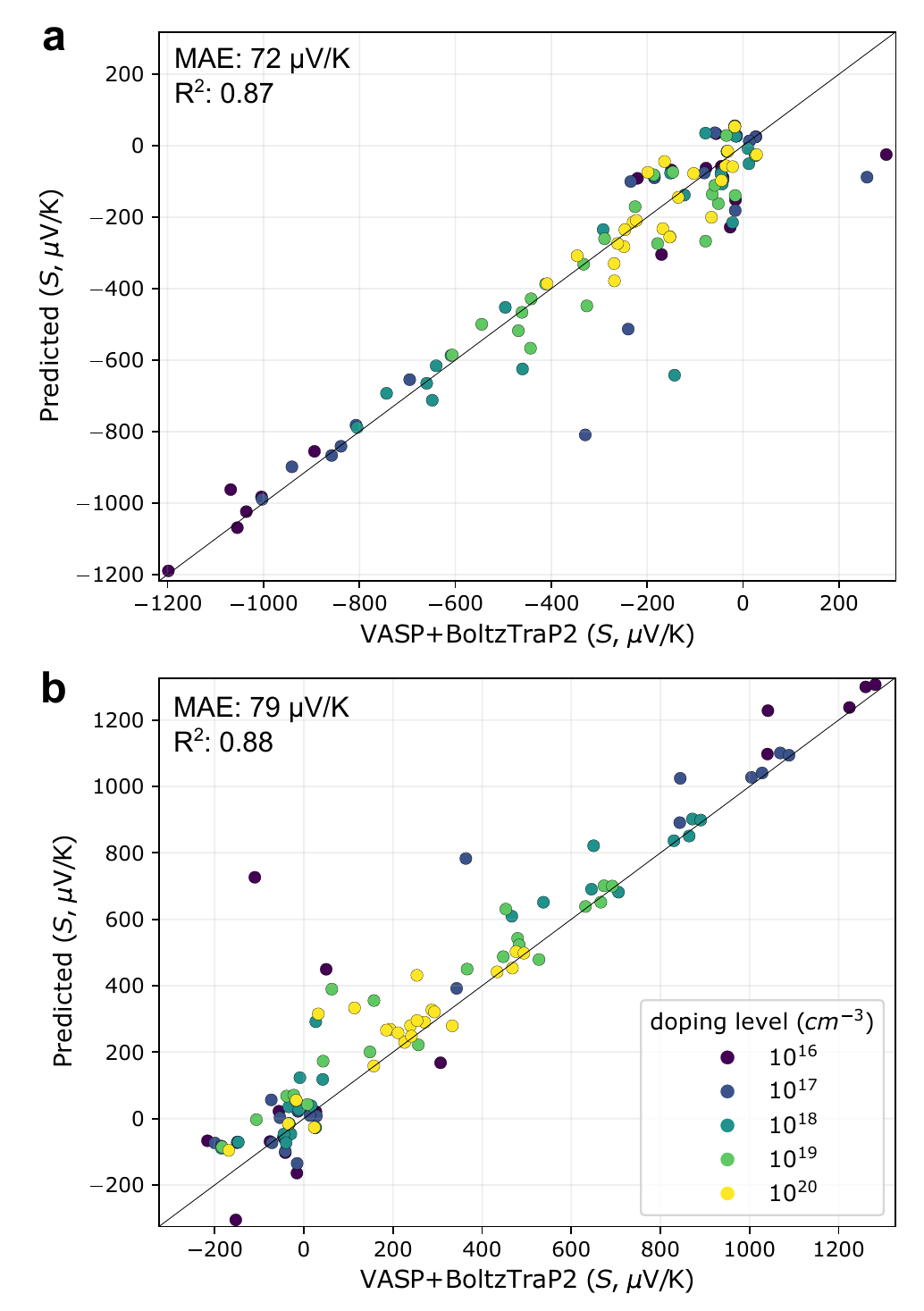}
	\caption{Seebeck coefficients at 700 K predicted with CraTENet+gap \textit{vs.} those computed using the \textit{ab initio} approach, for 23 Materials Projects compounds not found in the Ricci database, with \textbf{a)} $n$-type doping, and  \textbf{b)} $p$-type doping. Each point represents a particular compound at a particular doping level (e.g. \ce{SbTeIr} at $10^{20} \mathrm{cm}^{-3}$).}
	\label{fgr:vasp-vs-pred-mp}
\end{figure}

Making predictions for tens of thousands of compounds with the CraTENet model is computationally inexpensive in comparison with \textit{ab initio} calculations, since inference is fast, aided by the use of GPUs and the inherent parallelism in neural networks. After performing inference on this space, we selected 23 materials from this collection that spanned a range of different thermoelectric properties and band gaps. For example, the predicted Seebeck values ranged from -1200 to 1200 $\mu V/K$. We found that, when the band gap was included, the $R^2$ was between 0.87 and 0.88, and the MAE was between 72 and 79 $\mu V/K$, when comparing the values produced using the CraTENet+gap model and those obtained through \textit{ab initio} methods (Figure \ref{fgr:vasp-vs-pred-mp}).

Moreover, we extracted the top 1000 compounds by predicted power factor, for each of $p$ and $n$ doping types (the lists are provided in the dataset accompanying this article). We selected 4 $p$-type selenides for performing \textit{ab initio} calculations: \ce{GaCuTeSe}, \ce{InCuTeSe}, \ce{CeSbSe}, and \ce{Cu2SiSe3}. These compounds do not appear to have been studied as thermoelectrics before, but they seem promising as they include elements like Cu, In, Sb, and Te that are present in well-known thermoelectrics. After carrying out \textit{ab initio} calculations, we found generally good agreement with the CraTENet predictions (Figure \ref{fgr:vasp-vs-pred-Cu2SiSe3}; see Supplementary Figures 4-11 for more comprehensive plots of the predictions).

\begin{figure}[H]
    \centering
	\includegraphics[scale=0.55]{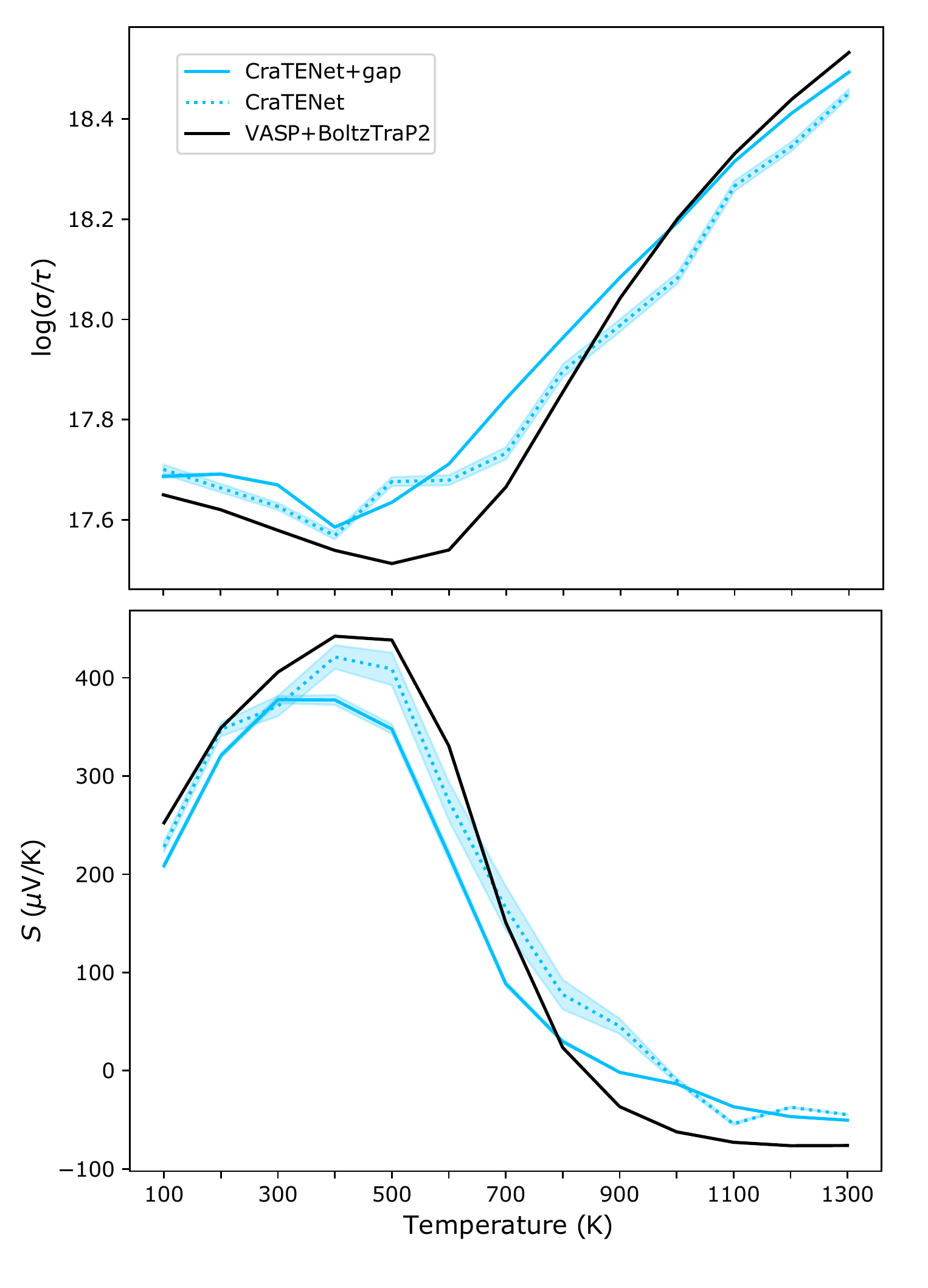}
	\caption{Predictions of the Seebeck and $\log \sigma$ for \ce{Cu2SiSe3} using the CraTENet models and the \textit{ab initio} procedure, for $p$-type doping, at a level of $10^{19}$ $\mathrm{cm}^{-3}$. The band gap value used, 0.242 eV, was obtained from the Materials Project. The shaded regions represent the $\pm$ standard deviation (\textit{i.e.} the square root of the predicted variance).}
	\label{fgr:vasp-vs-pred-Cu2SiSe3}
\end{figure}

\subsubsection{Hypothetical Selenides}

Since the CraTENet model requires only composition as input, it is conceivable that arbitrarily large hypothetical composition spaces could be generated and then processed by the model. SMACT is a software library that facilitates the generation of composition spaces, while adhering to chemical bonding rules, resulting in compositions which are chemically sensible \cite{davies2019smact}. Selenium-based materials are very promising thermoelectrics, because they exhibit similar properties as record-holding thermoelectric tellurides, but with the advantage that Se is much more Earth-abundant and cheaper than Te. We then chose to focus on creating a composition space of ternary selenides. Using SMACT, we generated 269,846 ternary selenide compositions, containing elements with an atomic number less than 84 (to avoid the heavy radioactive elements). The CraTENet and CraTENet+gap models were then used to make predictions of the thermoelectric transport properties of these compositions. As the CraTENet+gap model requires a band gap, we use our composition-only CrabNet band gap predictor as the source of the band gaps for this space. Since there is uncertainty in the band gap prediction, we make a separate prediction of thermoelectric transport properties using the predicted gap, the predicted gap plus the standard deviation, and the predicted gap minus the standard deviation. We find that this technique is useful for understanding the sensitivity of the predictions to the band gap value for a particular composition.

\begin{figure}[H]
    \centering
	\includegraphics[scale=0.42]{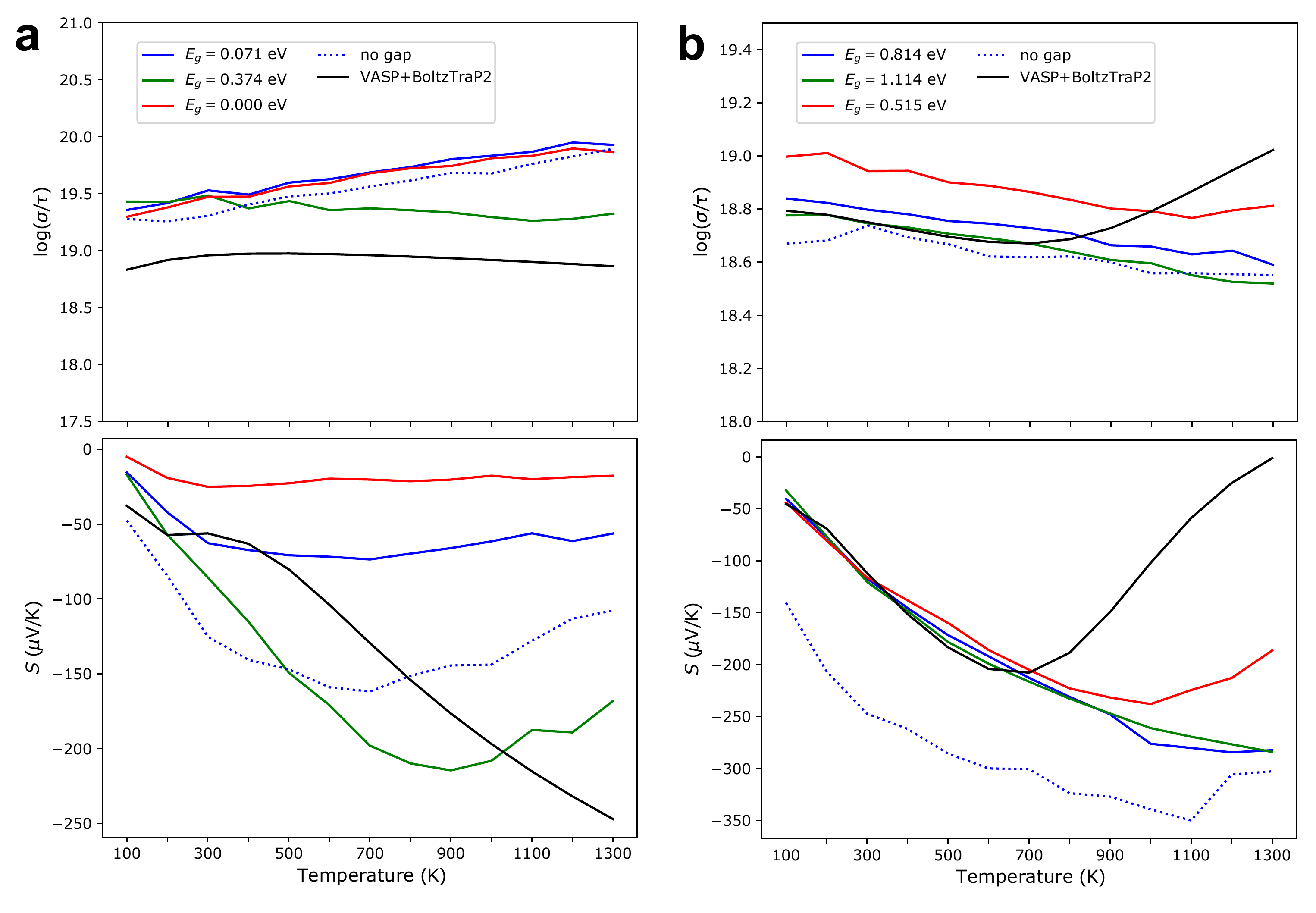}
	\caption{Predictions of the Seebeck and $\log \sigma$ for \textbf{a}) \ce{LiBiSe2}, and \textbf{b}) \ce{NaTlSe2} using the CraTENet models and the \textit{ab initio} procedure, for $n$-type doping, at a level of $10^{20}$ $\mathrm{cm}^{-3}$. Predicted band gap values were used: blue represents the initial prediction, green represents the prediction plus the predicted standard deviation, and red represents the prediction minus the predicted standard deviation (i.e. square root of the predicted variance).}
	\label{fgr:vasp-vs-pred-LiBiSe2-NaTlSe2}
\end{figure}

Having made predictions on these SMACT-generated selenides, we then rank the compositions by power factor (as described in the previous section). We make the top 1000 compositions publicly accessible in the code and dataset repository accompanying this article. There are several interesting selenides in that list, involving elements like bismuth (e.g. \ce{LiBiSe_2}) or thallium (e.g. \ce{NaTlSe_2}) which are often present in known thermoelectric materials.  To the best of our knowledge, these compounds have not been studied as thermoelectrics in the literature. To validate the model's predictions, we carried out \textit{ab initio} calculations on these two compounds, given that their structures are reported in the OQMD database \cite{saal2013materials}. A comparison of the predictions and the \textit{ab initio} values for each is provided in Figures \ref{fgr:vasp-vs-pred-LiBiSe2-NaTlSe2}a and \ref{fgr:vasp-vs-pred-LiBiSe2-NaTlSe2}b. (See also Supplementary Figures 12-15 for more comprehensive plots of the predictions). 

In the absence of DFT-calculated band gaps as input, the performance of the CraTENet model for these compounds is not as impressive in predicting the DFT-calculated values of the transport coefficients. The model using the predicted band gaps as an input seems to perform generally better than the model with no gap, but the deviations are still considerable, especially at high temperatures. All models, for example, overestimate the electrical conductivity of \ce{LiBiSe2} by at least half an order of magnitude. Still, the DFT calculations confirm, within their own limitations, that these compounds have attractive values of the electronic transport coefficients; they deserve further investigation, either using more accurate theoretical predictions with methods beyond the GGA and the CRTA, or experimentally. Clearly, the main use of the methods presented here cannot be the quantitative prediction of the transport properties of individual compounds, but rather the identification of interesting candidates in unexplored regions of the compositional space.

\section{Conclusions}

Approaches based on HTS combined with ML seem promising for suggesting novel candidate materials, since very large areas of chemical space can be examined quickly and efficiently. Here, we have shown that such an approach can be used to identify promising candidate thermoelectric materials based on the screening of potential compositions only, optionally supplemented with band gaps.

Several aspects of the approach described here contribute to its utility. First, the use of multi-output regression is helpful, and well-suited to the problem, since thermoelectric transport properties are dependent on factors such as temperature, doping level, and doping type. Conversely, an approach that requires parameters such as the temperature, doping level and doping type as input is problematic, since it increases the dimensionality of the input space, and also leads to inputs that resemble each other closely, as a result of the combinatorial nature of such a dataset \cite{zahrt2020cautionary}.

Second, we believe that regression is a more useful choice for this learning task when compared to classification, in the context of searching for new materials. Several existing studies have involved the training of classification models of thermoelectric properties \cite{gaultois2016perspective, lu2022artificial}. These classification approaches involve predicting whether a thermoelectric property is in a desired range, or above (or below) a specified threshold. We argue that regression models, such as ours, provide a level of increased utility via their finer-grained predictions, which is critical when sifting through many thousands of potential candidates. A binary classifier simply provides no convenient means of differentiating the candidates labelled as promising. Although there is room for improvement in the quality of the predictions made by our regression models, we find that at the current performance level, the approach is effective at surfacing promising candidates.

Third, the use of an attention-based model, in combination with the Robust L2 loss, both leads to superior performance and provides unique advantages. The learned attention weights provide an opportunity to interpret the predictions made for a composition \cite{wang2022crabnet}, and this could be a useful aspect of using the CraTENet model when analyzing individual materials (rather than in bulk, as we have focused on here). Additionally, the Robust L2 loss is especially useful in that it allows the model to learn to quantify the uncertainty arising from mapping the composition (and optionally band gap) to thermoelectric properties. This provides users with a quantitative measure of the certainty of a prediction.

Future work will involve follow-up investigations of the candidate materials proposed here, using more rigorous \textit{ab initio} methods. Should the candidates continue to appear promising, attempts may be made to synthesize the materials and measure their thermoelectric properties in the laboratory. In terms of the model itself, future work may involve augmenting the objective so that it takes into account the shape of the underlying manifold on which the multiple target values exist \cite{liu2009multi}. It is important to note that optimal thermoelectric transport properties are not the only criteria that establishes a material as a practical thermoelectric; other properties, such as dopability and stability, need to be considered. Thus, the computational discovery of novel thermoelectrics will be aided by the development of a suite of predictive models. 

It is clear that the approach we describe depends heavily on the quality of the data it is trained on. The Ricci database was derived using theoretical constraints such as the CRTA for solving the Boltzmann transport equation, and the GGA for the exchange correlation functionals, which have important limitations. However, the approach we describe here can continue to be used with future databases of computed thermoelectric properties that will be obtained with more accurate theoretical methods, with improved data quality.

Finally, to demonstrate the predictions made by the CraTENet model, we have deployed an internet-accessible web browser-based application, located at \\\href{https://thermopower.materialis.ai}{\color{blue}{https://thermopower.materialis.ai}}, that allows a user to submit a material's composition and (optionally) its band gap, and returns thermoelectric transport property predictions for the material, as made by the CraTENet model.

\section{Acknowledgements}
This work was partially supported by computational resource donations from Amazon Web Services through the AWS Activate program, obtained with assistance from the Communitech Hub. We are also grateful to the UK Materials and Molecular Modelling Hub for computational resources in the Young facility, which is partially funded by EPSRC (EP/P020194/1 and EP/T022213/1).  

\section{Data Availability}

The data that support the findings of this study are available as follows: \\
The Ricci database is publicly available online at:\\ \href{https://datadryad.org/stash/dataset/doi:10.5061/dryad.gn001}{\color{blue}{https://datadryad.org/stash/dataset/doi:10.5061/dryad.gn001}}. The Materials Project data that was used to train the band gap predictor and form a composition search space are publicly available online at: \href{https://materialsproject.org/}{\color{blue}{https://materialsproject.org/}}. The pre-trained SkipAtom embeddings that were used as input to the neural network models are located at: \href{https://github.com/lantunes/skipatom}{\color{blue}{https://github.com/lantunes/skipatom}}. The OQMD data that was used to provide structures for the SMACT-generated selenides are publicly available online at: \href{https://oqmd.org/}{\color{blue}{https://oqmd.org/}}.

\section{Code Availability}

The code with the CraTENet implementation, and for pre-processing the data and reproducing the experiments,
is open source, released under the MIT License. The code repository is accessible online, at:
\href{https://github.com/lantunes/CraTENet}{\color{blue}{https://github.com/lantunes/CraTENet}}.

\section{Author Contributions}

R.G.-C. conceived the project. L.M.A, K.T.B. and R.G.-C. designed the experiments. L.M.A. conceived and implemented the model, and performed the experiments. L.M.A drafted the manuscript. R.G.-C. and K.T.B. supervised and guided the project. All authors reviewed, edited and approved the manuscript.

\section{Competing Interests}

The authors declare no competing interests.

\bibliographystyle{naturemag}
\bibliography{references}

\end{document}


\maketitle

\section*{Supplementary Notes}

\subsection*{1. Descriptor Comparison}

We evaluated the use of various descriptors with the Random Forest and CraTENet models. To evaluate a model with a particular descriptor, we performed a 90-10 holdout experiment using the Ricci database, focusing only on the $p$-type Seebeck entries at 600K and $10^{20}$ $\mathrm{cm}^{-3}$ doping. For the Random Forest model, we evaluated two descriptors: the Meredig descriptor \cite{meredig2014combinatorial} and a sum-pooled 200-dimensional SkipAtom representation \cite{antunes2022distributed}. For the CraTENet model, we evaluated two atom descriptors: a SkipAtom 200-dimensional atom descriptor, and a binary descriptor described in the original CGCNN work \cite{xie2018crystal}. Each model is thus evaluated with a local descriptor (i.e. the Meredig material descriptor or the binary atom descriptor) and a distributed representation (i.e. the sum-pooled SkipAtom material representation or the SkipAtom atomic representation). The results are given in Supplementary Table \ref{tab:descr-comp}.

\subsection*{2. Robust L1 and Robust L2 Loss Comparison}

We evaluated the performance of the CraTENet model when trained to minimize either the Robust L1 loss or the Robust L2 loss. For this comparison, we chose to use a single-head output model that makes predictions across 13 different temperatures, for a fixed doping type and doping level. Specifically, we created models that predict either $S$ or $\log \sigma$, at 13 different temperatures, for either $n$- or $p$-type doping, at a level of $10^{20}$ $\mathrm{cm}^{-3}$. The results are given in Supplementary Table \ref{tab:robustl1-vs-robustl2}.

\section*{Supplementary Tables}

\begin{table}[H]
\centering
\caption {Results of 90-10 holdout experiments using the CraTENet+gap model, consisting of 1, 2, or 3 output heads. Identical train-test splits were used throughout. Each number represents either the MAE or $R^2$ on the held-out test set of the dataset formed from the Ricci database, across all temperatures, doping levels, and doping types. The 1-head entry refers to 3 separate models, each with a single output head, for each of $S$, $\log \sigma$, and $\log PF$. The 2-head model consisted of outputs only for $S$ and $\log \sigma$, and thus results for $\log PF$ are not reported. The 3-head model consisted of a separate output head for each of $S$, $\log \sigma$, and $\log PF$. Bold values represent the best result for a given metric and transport property.}
\begin{tabular}{ c c c| c c| c c } 
\hline
& \multicolumn{2}{c}{$S$} & \multicolumn{2}{c}{$\log \sigma$} & \multicolumn{2}{c}{$\log PF$} \\
\# output heads & MAE ($\mu$V/K) & $R^2$ & MAE & \bf $R^2$ & MAE & $R^2$  \\
\hline
1 & 51.595 & 0.960 & 0.269 & 0.964 & 0.384 & 0.727 \\
2 & 50.333 & 0.961 & 0.259 & \bf 0.968 & - & - \\
3 & \bf 49.718 & \bf 0.962 & \bf 0.257 & \bf 0.968 & \bf 0.372 & \bf 0.738 \\
\hline
\end{tabular}
\label{tab:single-vs-multihead} 
\end{table}

\begin{table}[H]
\centering
\caption {Mean relative predicted variance for the predictions of the various transport properties made by both the CraTENet and CraTENet+gap models, for the MP-excluding-Ricci dataset. Each model makes 54,816 $\times$ 130 predictions for a given transport property (i.e. for different temperatures, doping levels and doping types). Here, for each prediction, we take the associated predicted variance and divide it by the absolute value of the mean (i.e. the predicted value of the property), to obtain the relative predicted variance. The numbers in the table represent the mean of the relative predicted variances.}
\begin{tabular}{ c c c c } 
\hline
& $S$ & $\log \sigma$ & $\log PF$ \\
\hline
CraTENet & 4.8456 & 0.0075 & 0.0114 \\
CraTENet+gap & 0.0928 & 0.0003 & 0.0032 \\
\hline
\end{tabular}
\label{tab:mean-rel-var} 
\end{table}

\begin{table}[H]
\centering
\caption {Results of 90-10 holdout experiments for the $p$-type Seebeck entries at 600K and $10^{20}$ $\mathrm{cm}^{-3}$ doping of the Ricci database. Identical train-test splits were used throughout. Band gap is not provided to the models. The CraTENet model consisted of a single output head only. Bold results represent the best results for a model type.}
\begin{tabular}{ c l c c } 
\hline
Model & Descriptor & $R^2$ & MAE ($\mu$V/K) \\
\hline
Random Forest & Meredig & \bf 0.643 & \bf 74 \\
Random Forest & sum-pooled SkipAtom 200-dim & 0.551 & 92 \\
\hline
CraTENet & SkipAtom 200-dim & \bf 0.587 & \bf 69 \\
CraTENet & CGCNN binary atom vector & 0.556 & 71 \\
\hline
\end{tabular}
\label{tab:descr-comp} 
\end{table}

\begin{table}[H]
\centering
\caption {Results for CraTENet models with a single output head that produce predictions for 13 temperatures, for a given thermoelectric transport property, at a doping level of $10^{20}$ $\mathrm{cm}^{-3}$, and for a specific doping type. The models were trained using either the Robust L1 or Robust L2 loss. Each column represents either the MAE or the $R^2$ across all temperatures and across all members of the test set of a 90-10 holdout experiment. Identical train-test splits were used throughout. MAE values for tasks involving prediction of $S$ are in units of $\mu$V/K. Bold values represent the best result for a loss for a particular metric.}
\begin{tabular}{ l c c| c c} 
\hline
& \multicolumn{2}{c}{Robust L1} & \multicolumn{2}{c}{Robust L2} \\
Task & MAE & $R^2$ & MAE & $R^2$  \\
\hline
$p$-type $S$ & 67 & 0.592 & \bf 66 & \bf 0.612 \\
$n$-type $S$ & 64 & 0.478 & \bf 62 & \bf 0.503 \\
$p$-type $\log \sigma$ & 0.411 & 0.750 & \bf 0.402 & \bf 0.764 \\
$n$-type $\log \sigma$ & 0.388 & 0.711 & \bf 0.379 & \bf 0.723 \\
$p$-type $S$ + gap & 35 & 0.914 & 35 & 0.914 \\
$n$-type $S$ + gap & 38 & 0.831 & 38 & \bf 0.832 \\
$p$-type $\log \sigma$ + gap & \bf 0.272 & 0.896 & 0.275 & \bf 0.898 \\
$n$-type $\log \sigma$ + gap & 0.274 & 0.860 & \bf 0.269 & \bf 0.865 \\
\hline
\end{tabular}
\label{tab:robustl1-vs-robustl2} 
\end{table}

\section*{Supplementary Figures}

\begin{figure}[H]
    \centering
    \includegraphics[scale=0.7]{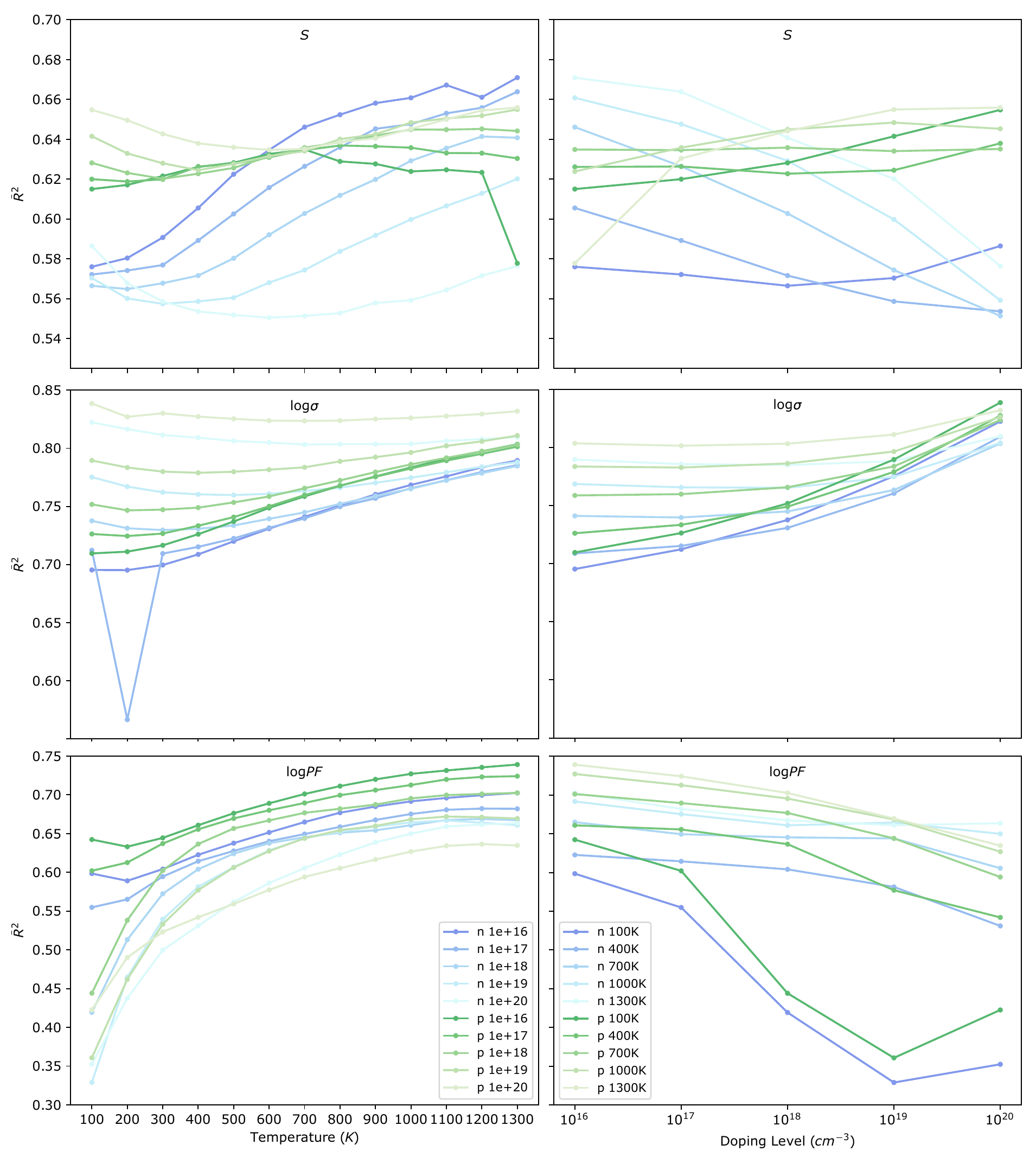}
	\caption{Plots of the agreement between the CraTENet and CraTENet+gap models, measured in terms of $R^2$, as a function of temperature and doping level, for each of the transport properties. Each point represents the average $R^2$ for the predictions made on the test set by the CraTENet and CraTENet+gap models, over the folds of 10-fold cross-validation experiments (each experiment utilized identical splits).}
	\label{fgr:gap-vs-no-gap-all}
\end{figure}

\begin{figure}[H]
    \centering
    \includegraphics[scale=0.9]{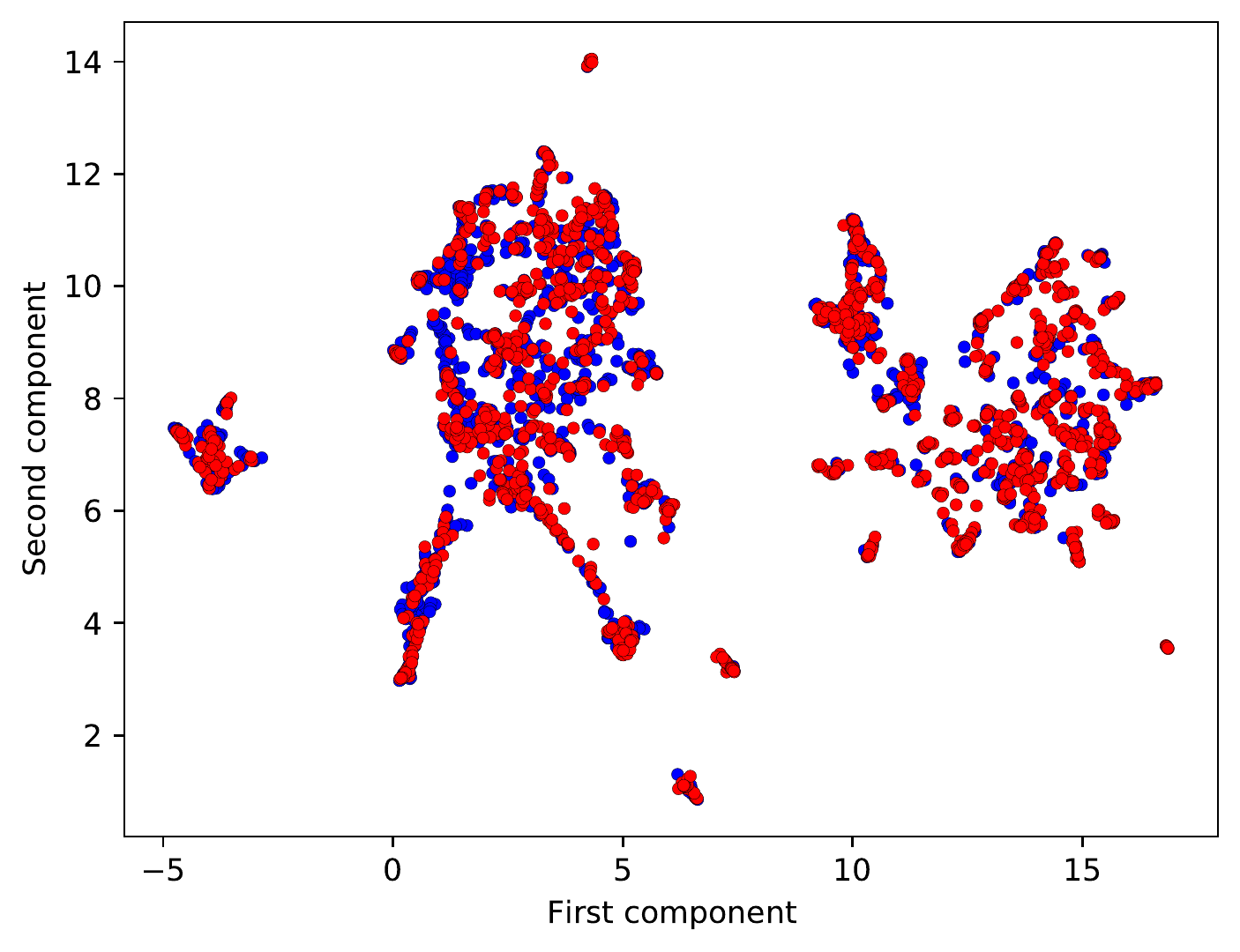}
	\caption{A plot of dimensionally-reduced 200-dimensional mean-pooled SkipAtom compound vectors for compounds from the Materials Project that are not in the Ricci database (red) and from the Ricci database (blue). UMAP \cite{mcinnes2018umap} with the Minkowski metric was used to reduce the compound vectors to 2 dimensions. The plot contains a random sampling of 3,000 compounds from each of the datasets. The plot supports the claim that the distributions of compounds in the datasets are similar.}
	\label{fgr:mp-not-in-db-vs-ricci}
\end{figure}

\begin{figure}[H]
    \centering
    \includegraphics[scale=0.75]{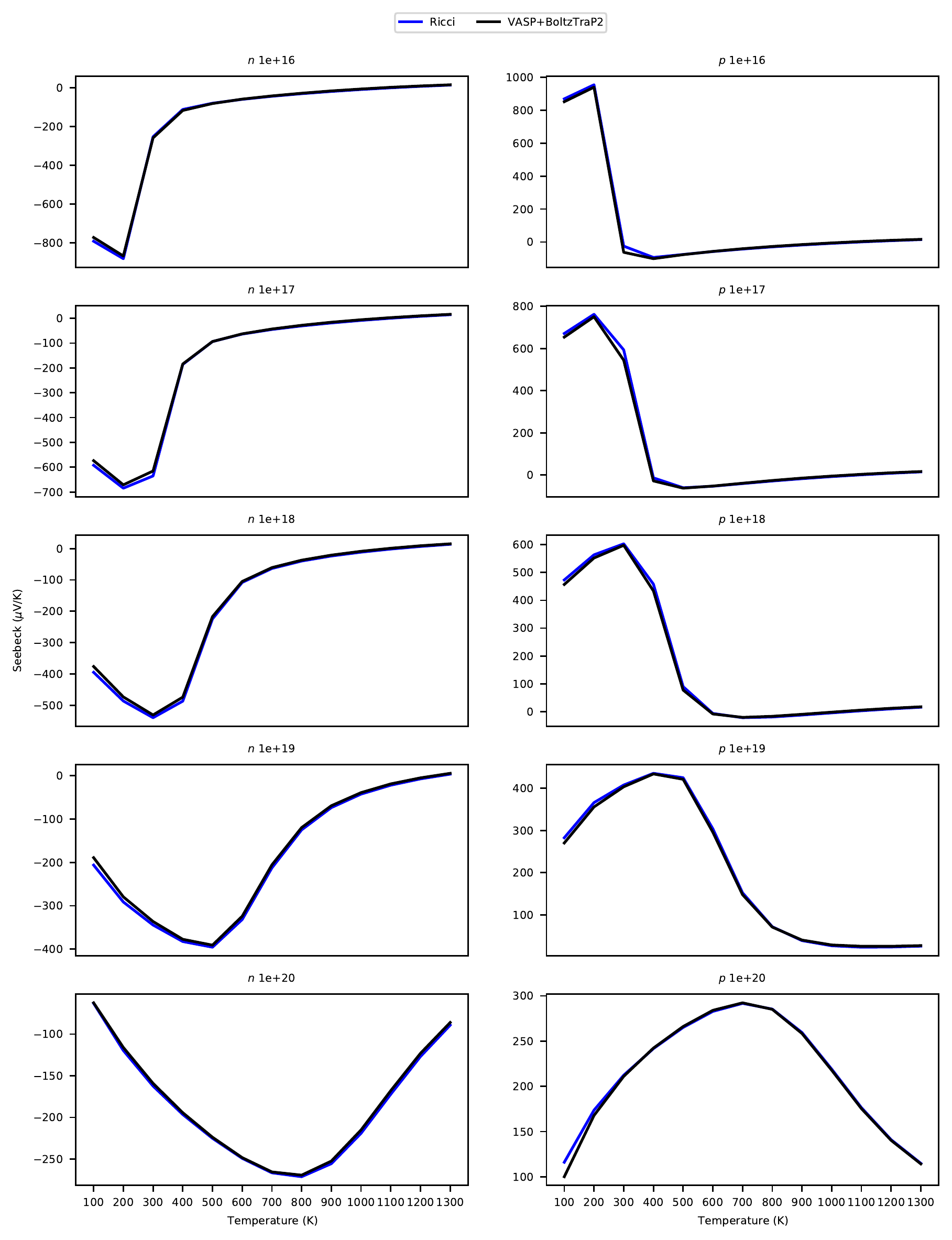}
	\caption{Plots comparing the Ricci database values for the Seebeck ($y$-axis, $\mu$V/K) to those produced by our \textit{ab initio} approach, for the compound HoSbPd (mp-567418). These results demonstrate that our \textit{ab initio} procedure emulates the one that was used to create the Ricci database. (Similar results were obtained for the electrical conductivity.)}
	\label{fgr:vasp-vs-actual}
\end{figure}

\begin{figure}[H]
    \centering
    \includegraphics[scale=0.75]{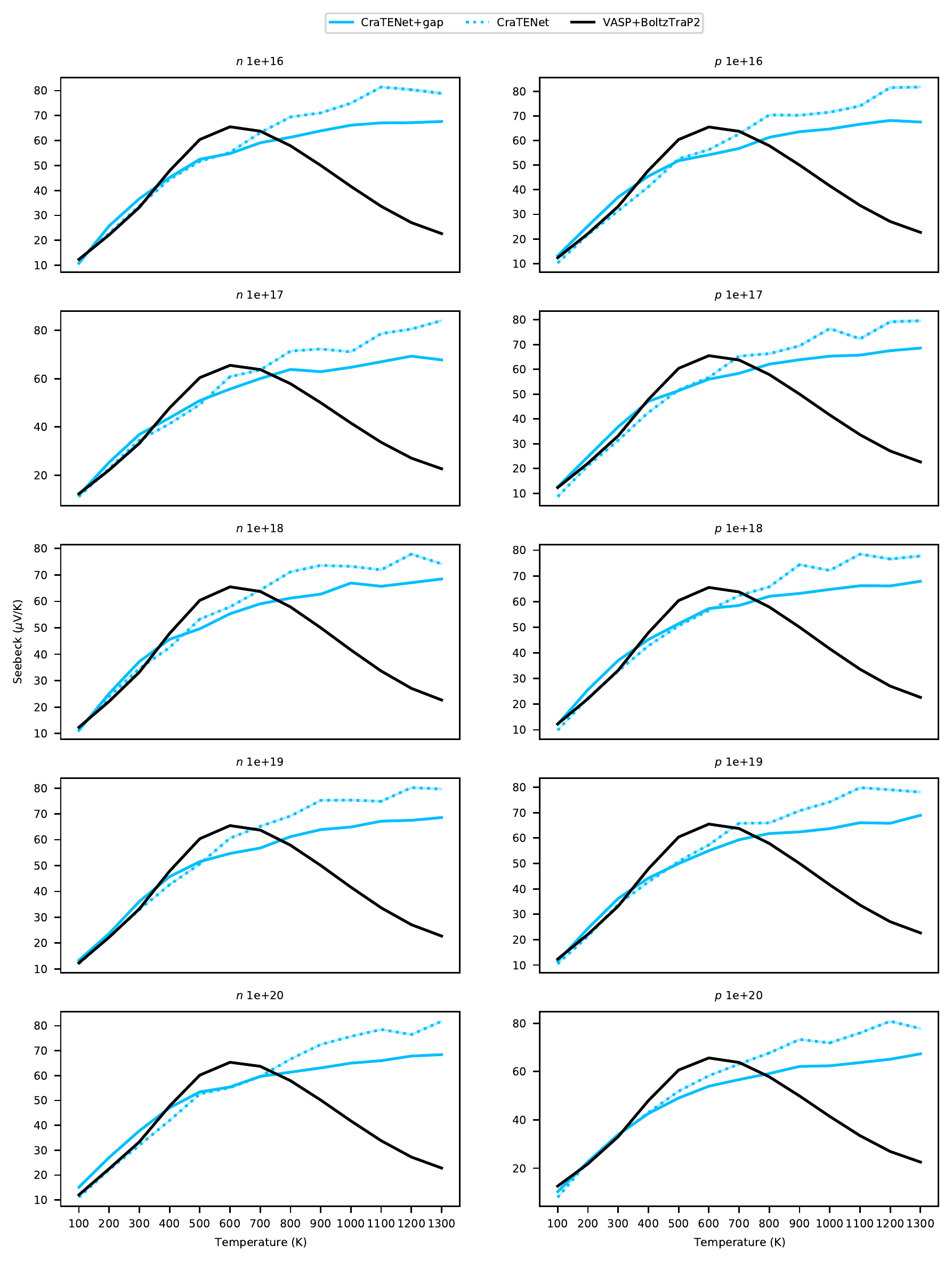}
	\caption{Plots of the Seebeck values for CeSbSe (mp-1103153) as predicted by the CraTENet models and by the \textit{ab initio} procedure. The band gap value used, 0.0 eV, was obtained from the Materials Project.}
	\label{fgr:vasp-vs-pred-seebeck-CeSbSe}
\end{figure}

\begin{figure}[H]
    \centering
    \includegraphics[scale=0.75]{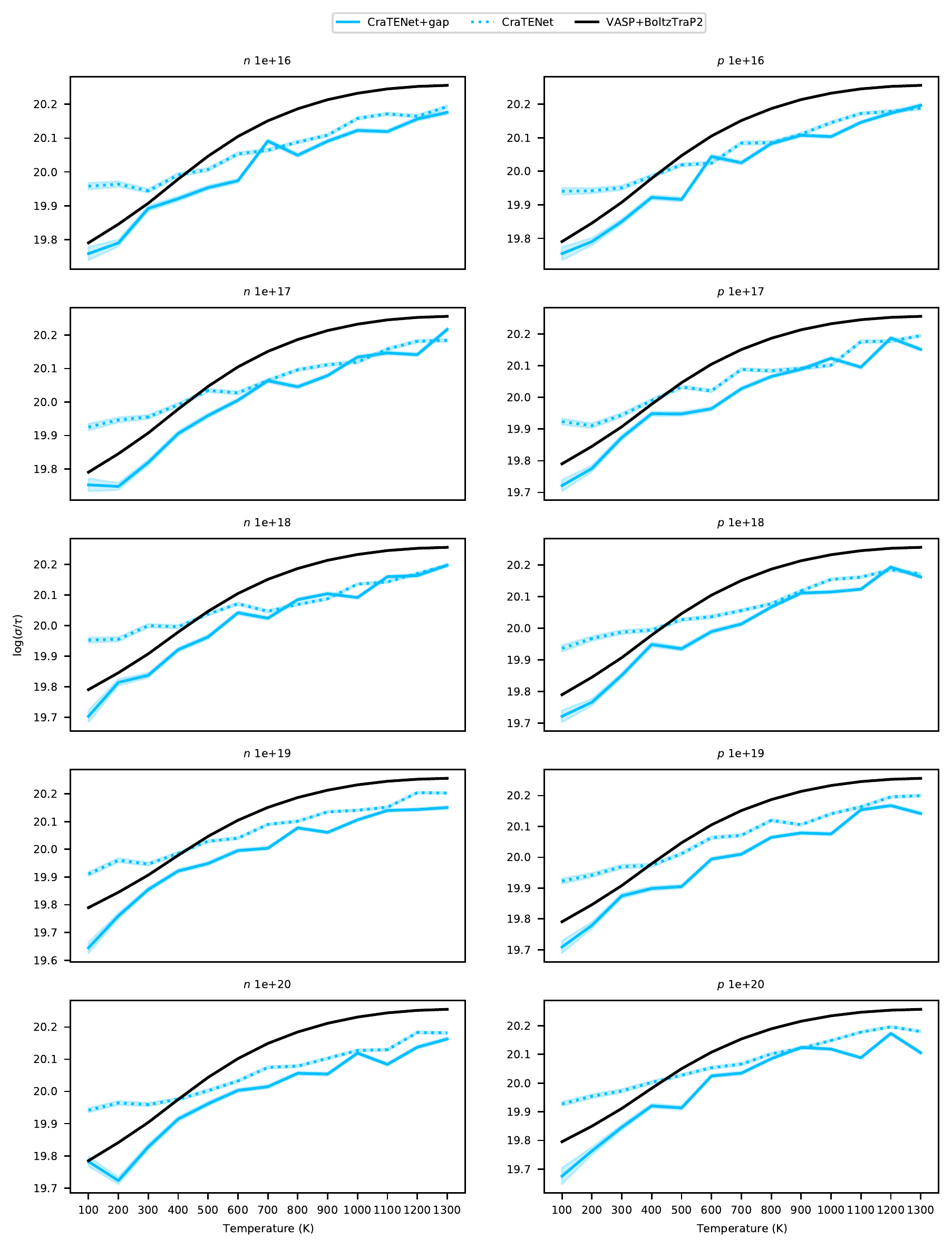}
	\caption{Plots of the $\log \sigma$ values for CeSbSe (mp-1103153) as predicted by the CraTENet models and by the \textit{ab initio} procedure. The band gap value used, 0.0 eV, was obtained from the Materials Project. The shaded regions represent the $\pm$ standard deviation (i.e. the square root of the predicted variance).}
	\label{fgr:vasp-vs-pred-log10cond-CeSbSe}
\end{figure}

\begin{figure}[H]
    \centering
    \includegraphics[scale=0.75]{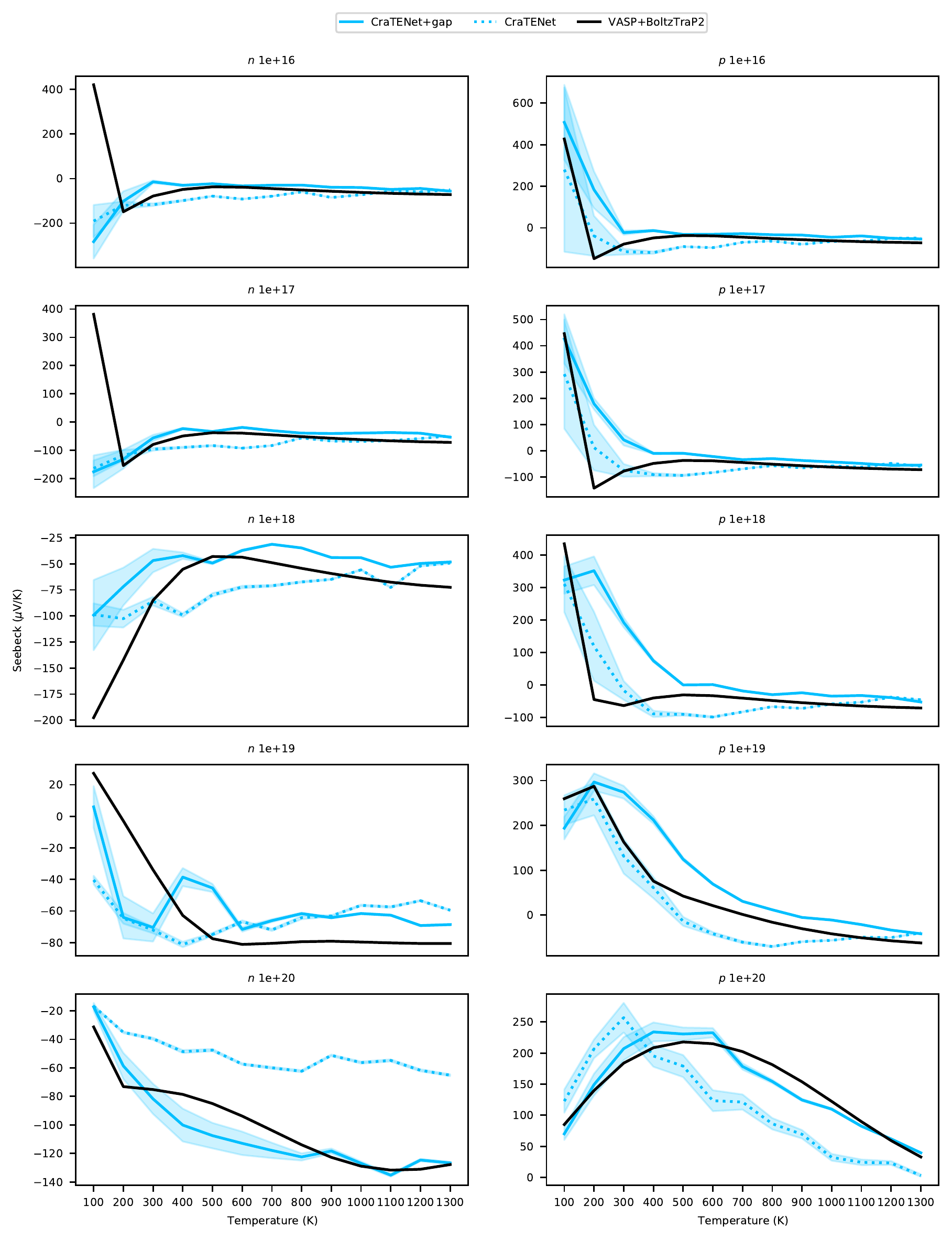}
	\caption{Plots of the Seebeck values for InCuTeSe (mp-1224187) as predicted by the CraTENet models and by the \textit{ab initio} procedure. The band gap value used, 0.164 eV, was obtained from the Materials Project. The shaded regions represent the $\pm$ standard deviation (i.e. the square root of the predicted variance).}
	\label{fgr:vasp-vs-pred-seebeck-InCuTeSe}
\end{figure}

\begin{figure}[H]
    \centering
    \includegraphics[scale=0.75]{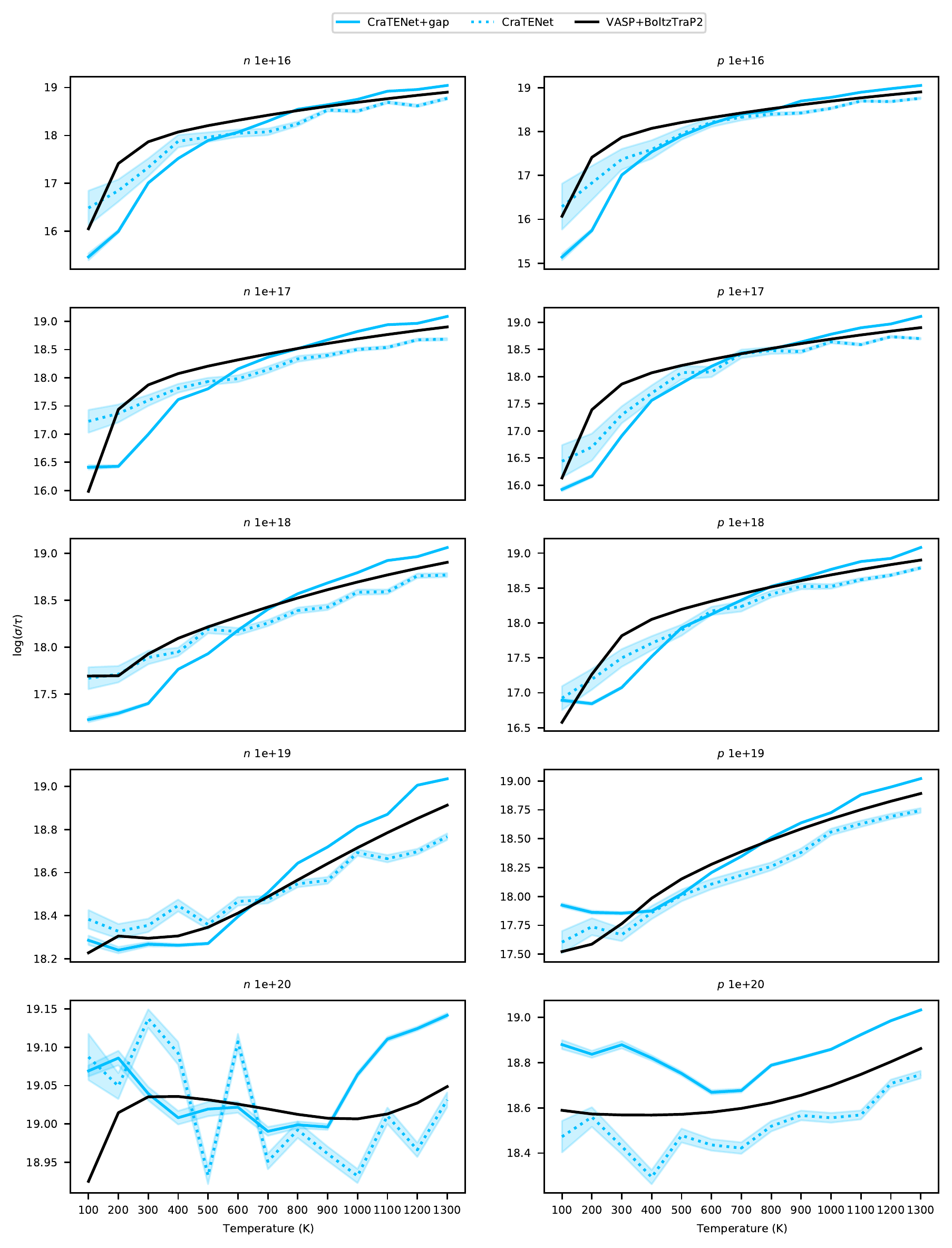}
	\caption{Plots of the $\log \sigma$ values for InCuTeSe (mp-1224187) as predicted by the CraTENet models and by the \textit{ab initio} procedure. The band gap value used, 0.164 eV, was obtained from the Materials Project. The shaded regions represent the $\pm$ standard deviation (i.e. the square root of the predicted variance).}
	\label{fgr:vasp-vs-pred-log10cond-InCuTeSe}
\end{figure}

\begin{figure}[H]
    \centering
    \includegraphics[scale=0.75]{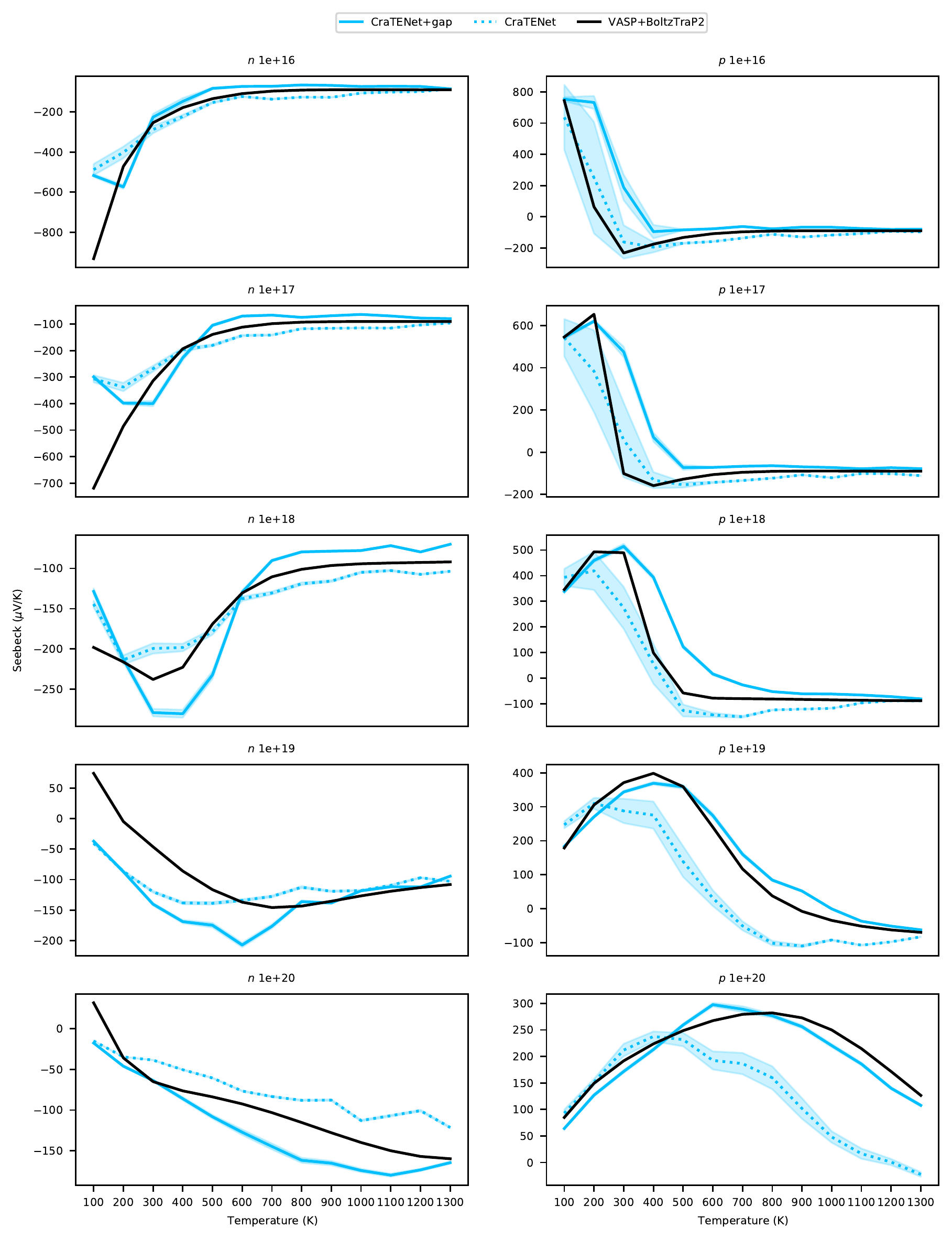}
	\caption{Plots of the Seebeck values for GaCuTeSe (mp-1224994) as predicted by the CraTENet models and by the \textit{ab initio} procedure. The band gap value used, 0.387 eV, was obtained from the Materials Project. The shaded regions represent the $\pm$ standard deviation (i.e. the square root of the predicted variance).}
	\label{fgr:vasp-vs-pred-seebeck-InCuTeSe}
\end{figure}

\begin{figure}[H]
    \centering
    \includegraphics[scale=0.75]{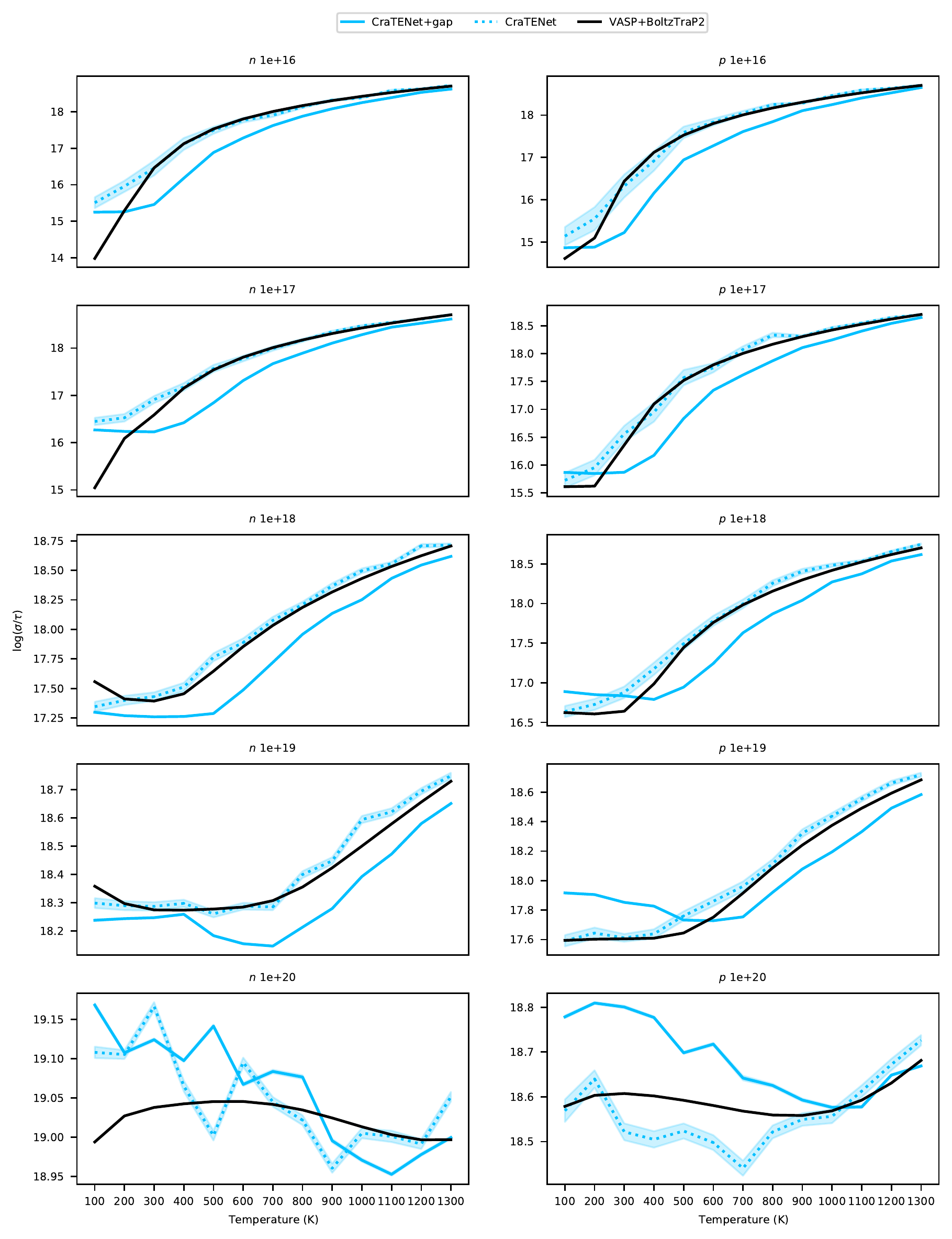}
	\caption{Plots of the $\log \sigma$ values for GaCuTeSe (mp-1224994) as predicted by the CraTENet models and by the \textit{ab initio} procedure. The band gap value used, 0.387 eV, was obtained from the Materials Project. The shaded regions represent the $\pm$ standard deviation (i.e. the square root of the predicted variance).}
	\label{fgr:vasp-vs-pred-log10cond-InCuTeSe}
\end{figure}

\begin{figure}[H]
    \centering
    \includegraphics[scale=0.75]{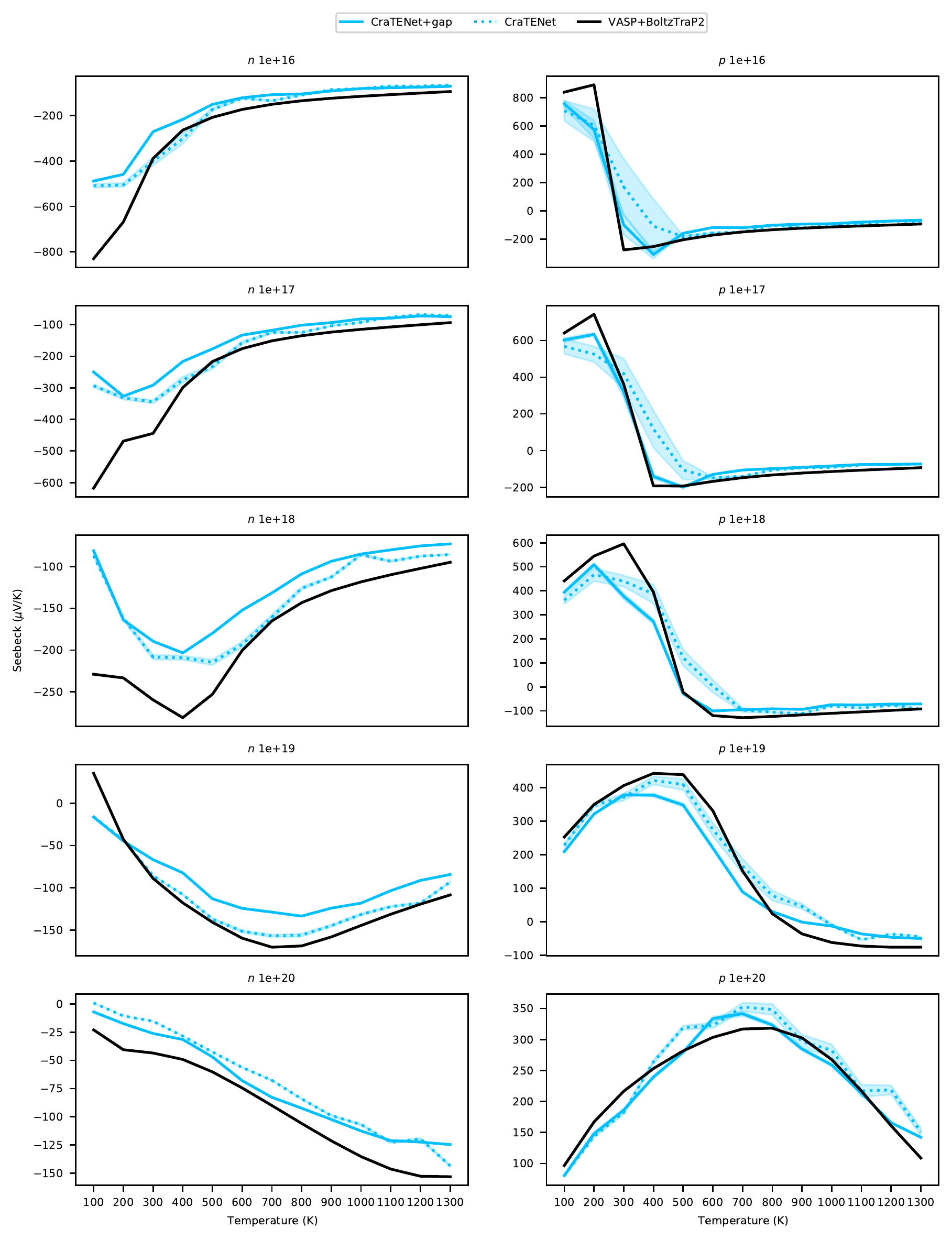}
	\caption{Plots of the Seebeck values for \ce{Cu2SiSe3} (mp-15896) as predicted by the CraTENet models and by the \textit{ab initio} procedure. The band gap value used, 0.242 eV, was obtained from the Materials Project. The shaded regions represent the $\pm$ standard deviation (i.e. the square root of the predicted variance).}
	\label{fgr:vasp-vs-pred-seebeck-Cu2SiSe3}
\end{figure}

\begin{figure}[H]
    \centering
    \includegraphics[scale=0.75]{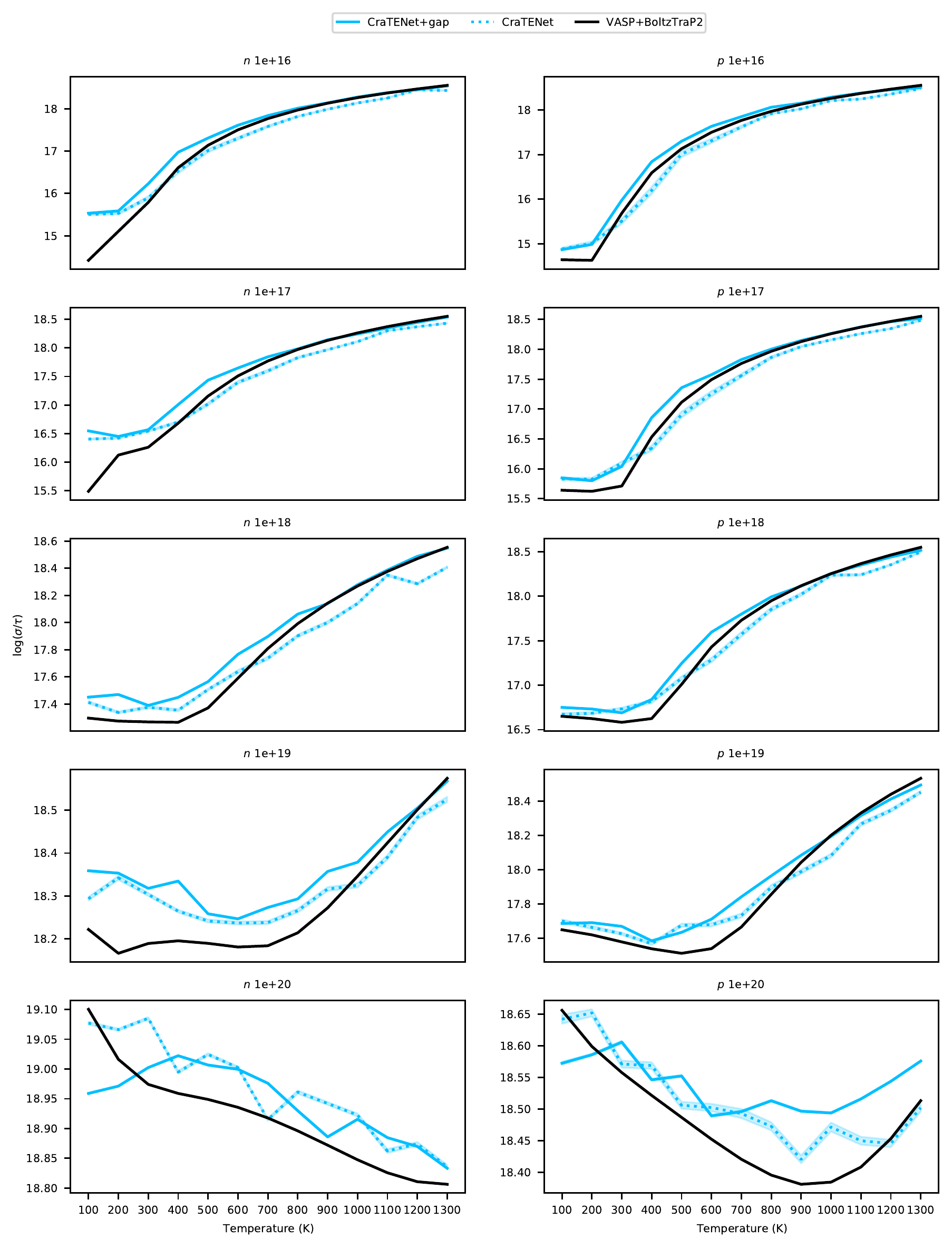}
	\caption{Plots of the $\log \sigma$ values for \ce{Cu2SiSe3} (mp-15896) as predicted by the CraTENet models and by the \textit{ab initio} procedure. The band gap value used, 0.242 eV, was obtained from the Materials Project. The shaded regions represent the $\pm$ standard deviation (i.e. the square root of the predicted variance).}
	\label{fgr:vasp-vs-pred-log10cond-Cu2SiSe3}
\end{figure}

\begin{figure}[H]
    \centering
    \includegraphics[scale=0.75]{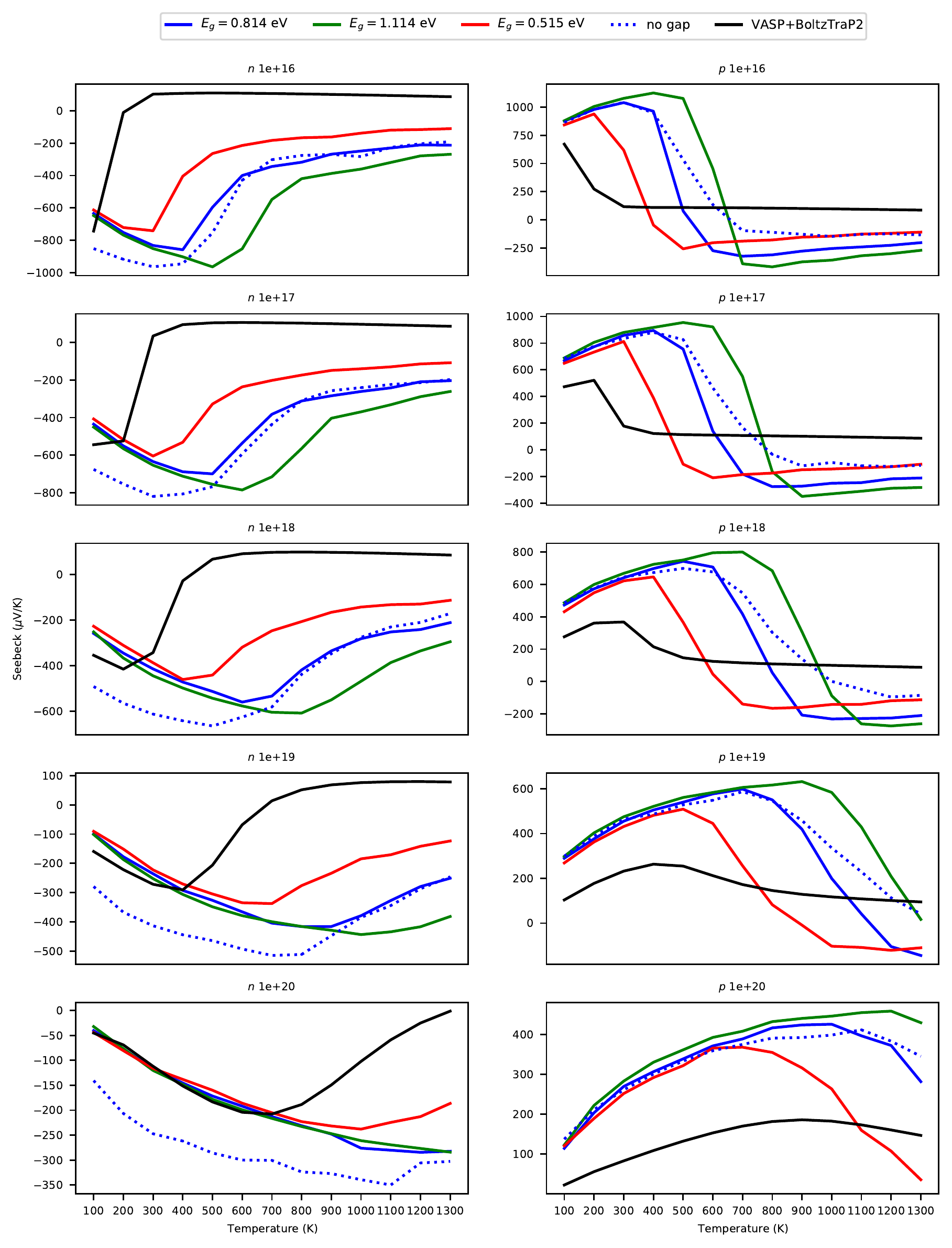}
	\caption{Plots of the Seebeck values for \ce{NaTlSe2} (oqmd-1482315) as predicted by the CraTENet models and by the \textit{ab initio} procedure. Predicted band gap values were used: blue represents the initial prediction, green represents the prediction plus the predicted standard deviation, and red represents the prediction minus the predicted standard deviation (i.e. square root of the predicted variance).}
	\label{fgr:vasp-vs-pred-seebeck-NaTlSe2}
\end{figure}

\begin{figure}[H]
    \centering
    \includegraphics[scale=0.75]{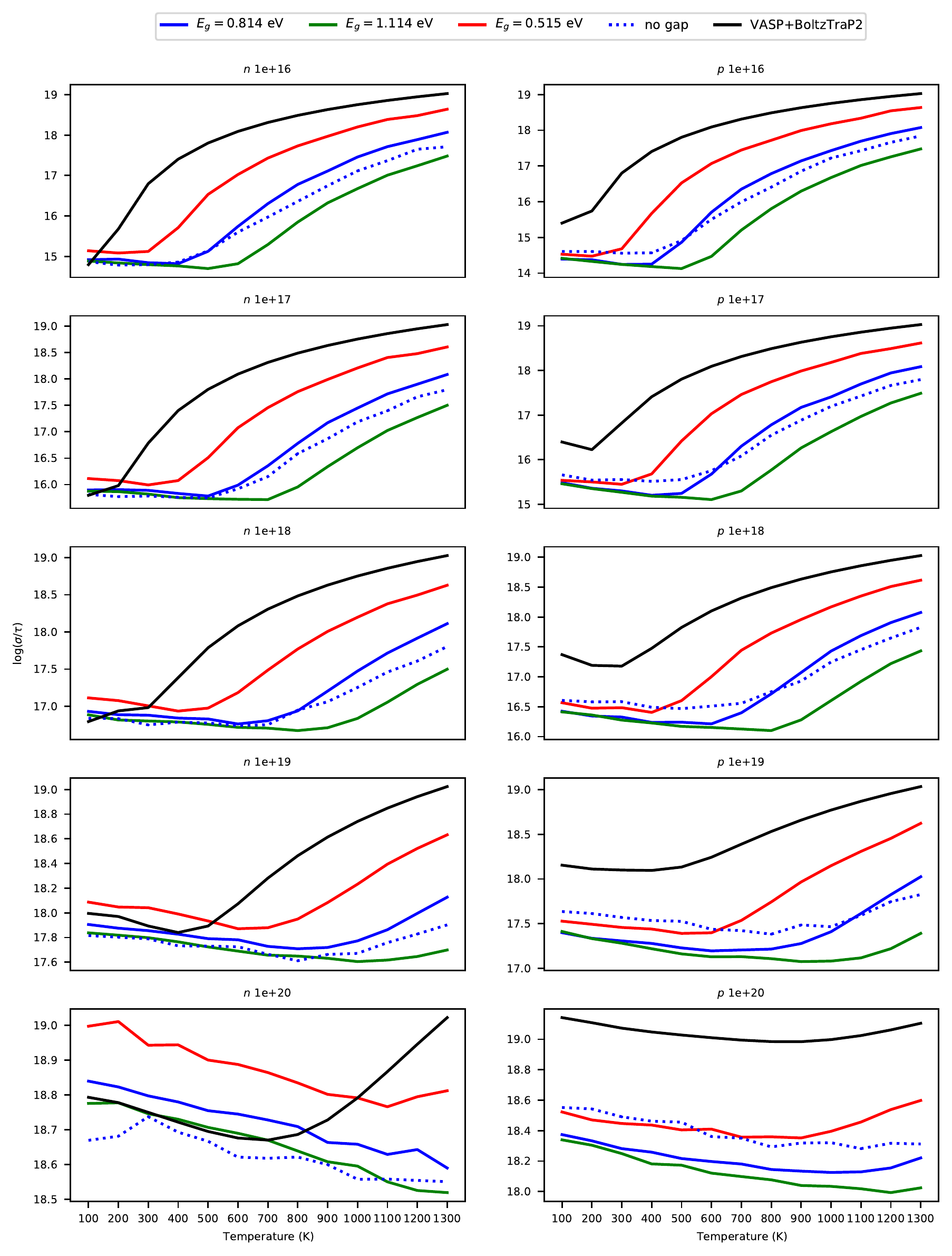}
	\caption{Plots of the $\log \sigma$ values for \ce{NaTlSe2} (oqmd-1482315) as predicted by the CraTENet models and by the \textit{ab initio} procedure. Predicted band gap values were used: blue represents the initial prediction, green represents the prediction plus the predicted standard deviation, and red represents the prediction minus the predicted standard deviation (i.e. square root of the predicted variance).}
	\label{fgr:vasp-vs-pred-log10cond-NaTlSe2}
\end{figure}

\begin{figure}[H]
    \centering
    \includegraphics[scale=0.75]{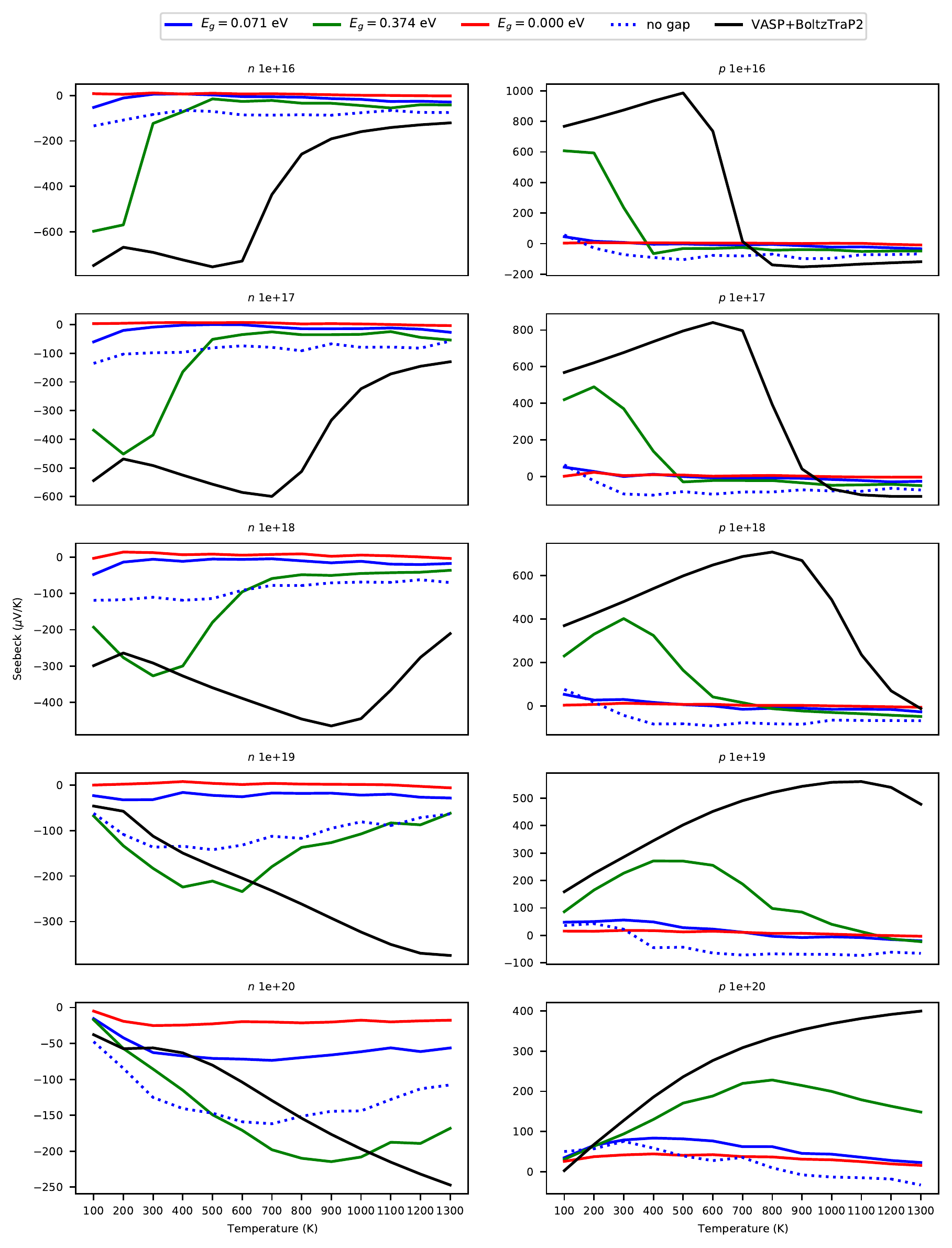}
	\caption{Plots of the Seebeck values for \ce{LiBiSe2} (oqmd-1442673) as predicted by the CraTENet models and by the \textit{ab initio} procedure. Predicted band gap values were used: blue represents the initial prediction, green represents the prediction plus the predicted standard deviation, and red represents the prediction minus the predicted standard deviation (i.e. square root of the predicted variance).}
	\label{fgr:vasp-vs-pred-seebeck-LiBiSe2}
\end{figure}

\begin{figure}[H]
    \centering
    \includegraphics[scale=0.75]{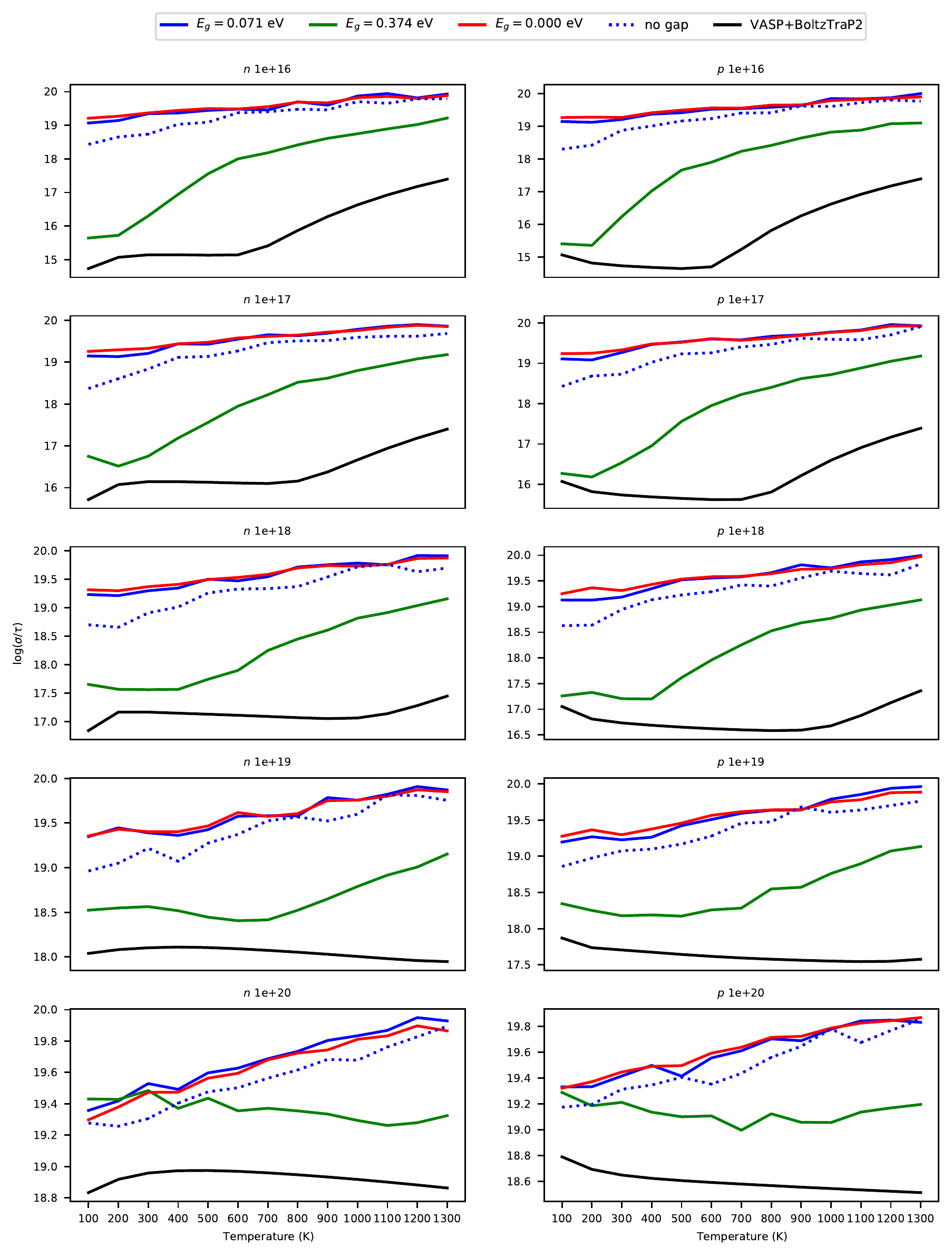}
	\caption{Plots of the $\log \sigma$ values for \ce{LiBiSe2} (oqmd-1442673) as predicted by the CraTENet models and by the \textit{ab initio} procedure. Predicted band gap values were used: blue represents the initial prediction, green represents the prediction plus the predicted standard deviation, and red represents the prediction minus the predicted standard deviation (i.e. square root of the predicted variance).}
	\label{fgr:vasp-vs-pred-log10cond-LiBiSe2}
\end{figure}

\bibliographystyle{naturemag}
\bibliography{references}